%% file: main.tex
\documentclass[acmsmall,screen,nonacm]{acmart}

\input{preamble}

\begin{document}

\title{DRReduce: Enhancing Syntax-Guided Program Reduction with Dependency Reconstruction}

\author{Qiong Feng}
\orcid{0000-0003-1667-8062}
\affiliation{%
  \institution{Nanjing University of Science and Technology}
  \city{Nanjing}
  \country{China}
}
\email{qiongfeng@njust.edu.cn}

\author{Xiaotian Ma}
\orcid{0009-0008-9121-3656}
\affiliation{%
  \institution{Nanjing University of Science and Technology}
  \city{Nanjing}
  \country{China}
}
\email{xyzboom@njust.edu.cn}

\author{Yongqiang Tian}
\orcid{0000-0003-1644-2965}
\affiliation{%
  \institution{Monash University}
  \city{Melbourne}
  \country{Australia}
}
\email{yongqiang.tian@monash.edu}

\author{Wei Song}
\orcid{0000-0002-4324-3382}
\affiliation{%
  \institution{Nanjing University of Science and Technology}
  \city{Nanjing}
  \country{China}
}
\email{wsong@njust.edu.cn}

\author{Peng Liang}
\orcid{0000-0002-2056-5346}
\affiliation{%
  \institution{School of Computer Science, Wuhan University}
  \city{Wuhan}
  \country{China}
}
\email{liangp@whu.edu.cn}

\renewcommand{\shortauthors}{Feng et al.}

\input{abstract}




\maketitle

\input{intro}

\input{motivation}
\input{approach}
\input{experiment-setup}
\input{results}

\input{discussion}

\input{related}
\input{conclusion}

\section*{Data Availability}\label{sec:DataAvailability}
Our data is publicly available at \url{https://github.com/XYZboom/DRReduceData}, which includes all bug-triggering programs, the programs reduced using different baselines, and the results of ablation experiments.

\section*{Acknowledgments}\label{sec:Acknowledgments}
This work has been partially supported by the National Natural Science Foundation of China (NSFC) with Grant No. 92582203.

\bibliographystyle{ACM-Reference-Format}
\bibliography{ref}

\end{document}

%% file: preamble.tex
\usepackage{diagbox}
\usepackage{algpseudocode}
\usepackage{booktabs}
\usepackage{adjustbox}
\usepackage[normalem]{ulem}
\usepackage{subcaption}
\usepackage{colortbl}
\usepackage[stable]{footmisc}
\usepackage{listings}
\usepackage{amsthm}

\usepackage{natbib}
\usepackage{algorithm}
\usepackage{algpseudocode}
\usepackage{syntax}
\usepackage{multirow}
\usepackage{simplebnf}
\usepackage{makecell}
\usepackage{array}
\usepackage{ragged2e}
\usepackage{threeparttable}
\usepackage{bbding}
\usepackage{rotating}
\usepackage{xcolor}

\lstdefinelanguage{Kotlin}{
  morekeywords={abstract, val, var, fun, if, else, while, for, in, return, open, class, interface, public, override},
  sensitive=true,
  morecomment=[l]{//},
  morecomment=[s]{/*}{*/},
  morestring=[b]",
  morestring=[s]{"""*}{*"""},
}
\lstdefinelanguage{Scala}{
  morekeywords={var, fun, if, else, while, for, in, return, open, class, interface, override, with, extends, c, abstract, def, return, public, default},
  sensitive=true,
  morecomment=[l]{//},
  morecomment=[s]{/*}{*/},
  morestring=[b]",
  morestring=[s]{"""*}{*"""},
}

\definecolor{codegreen}{rgb}{0,0.6,0}
\definecolor{codegray}{rgb}{0.5,0.5,0.5}
\definecolor{codepurple}{rgb}{0.58,0,0.82}
\definecolor{backcolour}{rgb}{0.95,0.95,0.92}
\definecolor{mylavender}{HTML}{C8BDE7}
\definecolor{softgreen}{HTML}{B2D1B9}
\definecolor{softpink}{HTML}{EBB7B6}
\definecolor{softpurple}{HTML}{8569CB}
\definecolor{softyellow}{HTML}{EBDEB7}

\usepackage{amsmath}
\usepackage{amsfonts}
\usepackage{balance}
\usepackage{enumitem}
\usepackage{graphicx}
\usepackage{url}
\usepackage{xspace}
\usepackage{tcolorbox}

\usepackage{tikz}
\usepackage{pgfplots}

\usetikzlibrary{shapes}
\usetikzlibrary{shapes.geometric}
\usetikzlibrary{arrows.meta, positioning}

\newcommand{\methodName}{\textsc{DRReduce}}
\newcommand{\toolName}{\textsc{DRReduce}}

\newcommand{\gccname}{\mbox{GCC}\xspace}
\newcommand{\llvmname}{\mbox{LLVM}\xspace}

\newcommand{\jdk}[1]{\href{https://bugs.openjdk.org/browse/JDK-#1?filter=allissues}{JDK-#1}}
\newcommand{\gcc}[1]{\href{https://gcc.gnu.org/bugzilla/show_bug.cgi?id=#1}{gcc-#1}}

\newcommand{\clang}[1]{\href{https://bugs.llvm.org/show_bug.cgi?id=#1}{clang-#1}}
\newcommand{\cf}[1]{\href{https://github.com/typetools/checker-framework/issues/#1}{cf-#1}}

\newcommand{\ecj}[1]{\href{https://bugs.eclipse.org/bugs/show_bug.cgi?id=#1}{ecj-#1}}

\definecolor{BlindColorTolOne}{HTML}{332288}
\definecolor{BlindColorTolTwo}{HTML}{117733} 
\definecolor{BlindColorTolThree}{HTML}{44AA99}
\definecolor{BlindColorTolFour}{HTML}{88CCEE}
\definecolor{BlindColorTolFive}{HTML}{DDCC77}
\definecolor{BlindColorTolSix}{HTML}{CC6677} 
\definecolor{BlindColorTolSeven}{HTML}{AA4499}
\definecolor{BlindColorTolEight}{HTML}{882255}

\definecolor{BlindColorWongOne}{HTML}{000000} 
\definecolor{BlindColorWongTwo}{HTML}{E69F00}
\definecolor{BlindColorWongThree}{HTML}{56B4E9}
\definecolor{BlindColorWongFour}{HTML}{009E73}
\definecolor{BlindColorWongFive}{HTML}{F0E442}
\definecolor{BlindColorWongSix}{HTML}{0072B2} 
\definecolor{BlindColorWongSeven}{HTML}{D55E00}
\definecolor{BlindColorWongEight}{HTML}{CC79A7}

\definecolor{mygreen}{HTML}{02818a}

\mathchardef\mhyphen="2D

\newcounter{FindingCounter}

\newcounter{myUniqueIdCounter}

\makeatletter
\newcommand{\myGetOrAssignID}[1]{%
  \ifcsname myMap@#1\endcsname%
    \csname myMap@#1\endcsname%
  \else%
    \stepcounter{myUniqueIdCounter}%
    \expandafter\xdef\csname myMap@#1\endcsname{\themyUniqueIdCounter}%
    \themyUniqueIdCounter%
  \fi%
}
\makeatother

\newcommand{\anonymizedId}[1]{\ifx\useAnonymizedId\undefined%
  #1%
\else%
  \myGetOrAssignID{#1}%
\fi%
}

\usepackage{cleveref} 

\Crefname{algocf}{Algorithm}{Algorithms}
\crefname{algocf}{Algorithm}{Algorithms}

\Crefname{algorithm}{Algorithm}{Algorithms}
\crefname{algorithm}{Algorithm}{Algorithms}

\crefname{appendix}{Appendix}{Appendices}
\Crefname{appendix}{Appendix}{Appendices}

\Crefname{figure}{Figure}{Figures}
\crefname{figure}{Figure}{Figures}

\crefname{listing}{Listing}{Listings}
\Crefname{listing}{Listing}{Listings}

\Crefname{table}{Table}{Tables}
\crefname{table}{Table}{Tables}

\crefname{thm}{Theorem}{Theorems}
\Crefname{thm}{Theorem}{Theorems}

\crefname{equation}{Equation}{Equations}
\Crefname{equation}{Equation}{Equations}

\crefformat{chapter}{\S~#2#1#3}
\crefmultiformat{chapter}{\S\S~#2#1#3}{ and~#2#1#3}{, #2#1#3}{, and~#2#1#3}

\crefformat{section}{\S~#2#1#3}
\crefmultiformat{section}{\S\S~#2#1#3}{ and~#2#1#3}{, #2#1#3}{, and~#2#1#3}

%% file: abstract.tex
\begin{abstract}

Program reduction is a critical technique for simplifying large, failure-inducing programs into minimal reproducible test cases. Language-specific tools such as CReduce achieve strong performance by leveraging deep semantic knowledge of C/C++, but are tightly coupled to a single language family. Language-agnostic reducers such as Perses address this by applying syntax-guided search across any grammar, yet share a fundamental limitation: deleting a node or subtree in isolation often breaks semantic coherence --- such as leaving unresolved references or inconsistent signatures --- causing the property checker to reject the deletion and forcing the reducer to backtrack, limiting overall reduction effectiveness and efficiency.

In this paper, we propose \toolName{}, a framework that bridges this gap by augmenting language-agnostic syntactic reduction with a lightweight semantic layer: \textit{dependency reconstruction}, which repairs program dependencies broken by a deletion in order to preserve the semantic validity of intermediate programs and increase the acceptance rate of the property checker. \toolName{} constructs a semantic dependency graph from the input program, performs semantically coherent deletions with dependency reconstruction, and delegates further minimization to a syntax-guided reducer.

We implement \toolName{} for C and Java and evaluate it on real-world bug-triggering programs. Compared to state-of-the-art syntax-guided reducers, \toolName{} achieves average size reductions of 51.9\%, 14.9\%, and 19.8\% over Perses, WDD, and CDD respectively, while completing reduction faster on the majority of programs. Compared to language-specific tools, \toolName{} achieves results comparable to CReduce and Latra without any language-specific transformation rules, at 3.3$\times$ and 1.2$\times$ higher efficiency than CReduce and Latra on average, respectively. An ablation study confirms that dependency reconstruction reduces query invocations by 80.2\%, reduction time by 58.7\%, and final token count by over 55.1\%.

\end{abstract}

%% file: intro.tex
\section{Introduction}
\label{sec:intro}

Program reduction is a critical debugging technique that simplifies large, failure-inducing programs into minimal reproducible test cases~\cite{yangfinding2011,sunfinding2016,lidbury-many-core-2015}. A minimal test case is easier to inspect, faster to replay, and more likely to isolate the root cause of a bug. As software systems grow in complexity and automated testing generates increasingly large failure-inducing inputs, the ability to reduce programs quickly and effectively has become essential to the debugging workflow of compiler developers and language implementors~\cite{ddmin,misherghi2006hdd,regehr2012test,zhang2021sanrazor,perses-trec,perses-cdd,zhang2024lpr,perses-sfc}. For instance, both \gccname{} and \llvmname{} explicitly require bug-triggering test cases to be minimized to reduce the workload of developers and increase the bug fix probabilities~\cite{gcc-bugs-guide,llvm-how-to-submit-bug}.

\textit{Reduction with language-specific transformations.}
The most effective program reducers leverage deep knowledge of a single target language. For example, CReduce~\cite{regehr2012test} targets C/C++ programs and achieves remarkably small outputs by applying a rich set of semantics-aware transformations --- inlining functions, simplifying types, removing unused declarations --- that are carefully engineered to preserve compilability and bug-triggering behavior. Similar tools exist for specific domains: llvm-reduce for LLVM IR~\cite{trevino2019llvm-reduce}. More recently, Latra~\cite{xu2025latra} has reduced the engineering burden of building such reducers by allowing users to express transformations as match-rewrite template pairs in a domain-specific language; for C, Latra implements 27 such rules and achieves results statistically comparable to CReduce. These tools work well precisely because they encode deep knowledge of a single language's semantics. However, this strength is also their limitation: the transformations are tightly coupled to the target language, and building a similar tool for another language requires substantial engineering effort.

\textit{Language-agnostic reduction.}
To address this, language-agnostic reducers such as Perses~\cite{sun2018perses}, WDD~\cite{zhou2024wdd} and ProbDD~\cite{pdd} apply syntax-guided search across any context-free grammar. Given a grammar and a property checker, these tools systematically attempt to delete or replace subtrees while preserving the property of interest. Because of operating on parse trees rather than semantic models, they require no language-specific knowledge and generalize immediately to any language with a formal grammar. In practice, they achieve competitive reduction ratios across a wide range of languages, making them the default choice when no dedicated reducer exists.

\textit{The semantic coherence problem.}
Despite their generality, syntax-only reducers share a fundamental limitation rooted in the mismatch between syntactic structure and semantic validity. Deleting a node or subtree in isolation frequently breaks semantic coherence: a removed function declaration leaves call sites with unresolved references, a deleted parameter creates a signature mismatch at every call site, and a removed type definition invalidates all variables of that type. The resulting intermediate program fails to compile, and the property checker rejects the deletion --- not because the reduction was semantically wrong, but because the surrounding program was left in an incoherent state. This forces the reducer to backtrack and try smaller subsets, multiplying query invocations and reducing overall efficiency. In effect, syntax-guided reducers pay the full search cost but cannot exploit semantic structure to recover rejected deletions.

\textbf{Our approach.}
In this paper, we propose \toolName{}, a framework that bridges this gap by augmenting language-agnostic syntactic reduction with a lightweight semantic layer. The key insight is that most semantic coherence failures following a deletion are \textit{local and repairable}.
When a node is deleted, other nodes that depended on it can be reconstructed — replaced with a default value of the appropriate type.
This reconstruction restores compilability without affecting whether the bug-triggering property is preserved.
We formalize this operation as \textit{dependency reconstruction} and integrate this operation into a three-stage reduction pipeline: \toolName{} first constructs a semantic dependency graph from the input programs, then applies dependency reconstruction to perform semantically coherent deletions, and finally delegates further minimization to a syntax-guided reducer such as Perses.

Critically, \toolName{} achieves this without language-specific transformation rules. The dependency graph is constructed from definition-use relationships that can be extracted from any language with a type system, and dependency reconstruction applies a uniform default-value replacement strategy that requires only type information. This makes \toolName{} applicable to any language for which a parser and basic semantic analysis are available.


To validate this generality and quantify the benefits of dependency reconstruction, we implement \toolName{} for C and Java and evaluate it on 28 real-world bug-triggering programs. Compared to state-of-the-art syntax-guided reducers (Perses, WDD, and CDD), \toolName{} reduces programs to 51.9\%, 14.9\%, and 19.8\% smaller sizes on average, while completing reduction faster on the majority of programs. Compared to reducers that incorporate language-specific transformations (CReduce and Latra), \toolName{} closely matches their effectiveness without using any language-specific transformation rules, at 3.3$\times$ and 1.2$\times$ higher efficiency on average, respectively.
An ablation study confirms that dependency reconstruction reduces query invocations by 80.2\%, reduction time by 58.7\%, and final token count by over 55.1\%.

This paper makes the following \textbf{contributions}:
\begin{itemize}
\item We identify semantic coherence breakage as a key source of inefficiency in language-agnostic program reduction, and formalize it through the notions of semantic dependency and dependency reconstruction (Section \ref{sec:approach-select} and Section \ref{sec:approach-reconstruct}).

\item We present \toolName{}, a language-agnostic reduction framework that integrates dependency reconstruction with syntax-guided reduction (Section \ref{sec:approach}).

\item We implement \toolName{} for two targets --- C and Java source code --- demonstrating its applicability across different source languages (Section \ref{sec:impl}).

\item We evaluate \toolName{} on real-world bug-triggering programs and show that it consistently outperforms state-of-the-art syntax-guided reducers in both effectiveness and efficiency (Section \ref{subsec:rq1}), while achieving results comparable to language-specific, semantic-transformation-based tools at substantially higher efficiency (Section \ref{subsec:rq2}).
\end{itemize}

The remainder of the paper is organized as follows: Section \ref{sec:motivation} presents a motivating example to illustrate the semantic coherence challenges that \toolName{} addresses. Section \ref{sec:approach} presents how \toolName{} is implemented. Section \ref{sec:experiment} and Section \ref{sec:results} detail the experimental setup and evaluation. Section~\ref{sec:discussion} discusses the limitations and future potential of \toolName{}, and Section~\ref{sec:threats} addresses threats to validity. Section \ref{sec:related} surveys related work, and Section \ref{sec:conclusion} concludes the paper.

%% file: motivation.tex
\section{A Motivating Example}
\label{sec:motivation}

\begin{figure}[htpb]
    \begin{subfigure}[t]{0.48\linewidth}
        \centering
        \includegraphics[width=\linewidth]{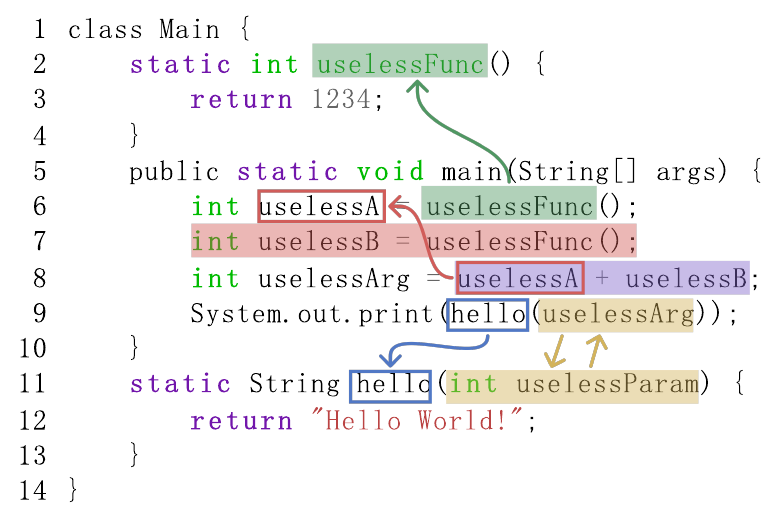}
        \caption{Before reduction. Arrows indicate direct semantic dependencies.}
        \label{fig:mtv-hello-before}
    \end{subfigure}
    \hfill
    \begin{subfigure}[t]{0.48\linewidth}
        \centering
        \includegraphics[width=\linewidth]{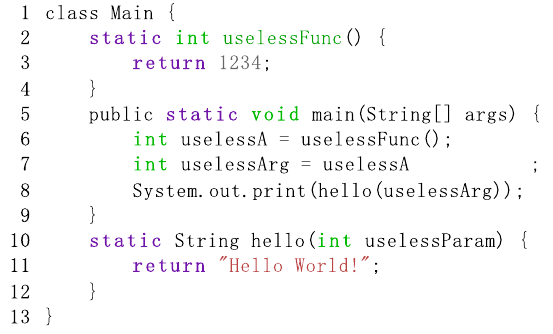}
        \caption{Reduction result generated by WDD.}
        \label{fig:mtv-hello-after}
    \end{subfigure}
    \caption{A Java program and its reduction result by WDD.}
    \label{fig:mtv-hello}
\end{figure}

\begin{figure}
    \centering
    \includegraphics[width=0.65\linewidth]{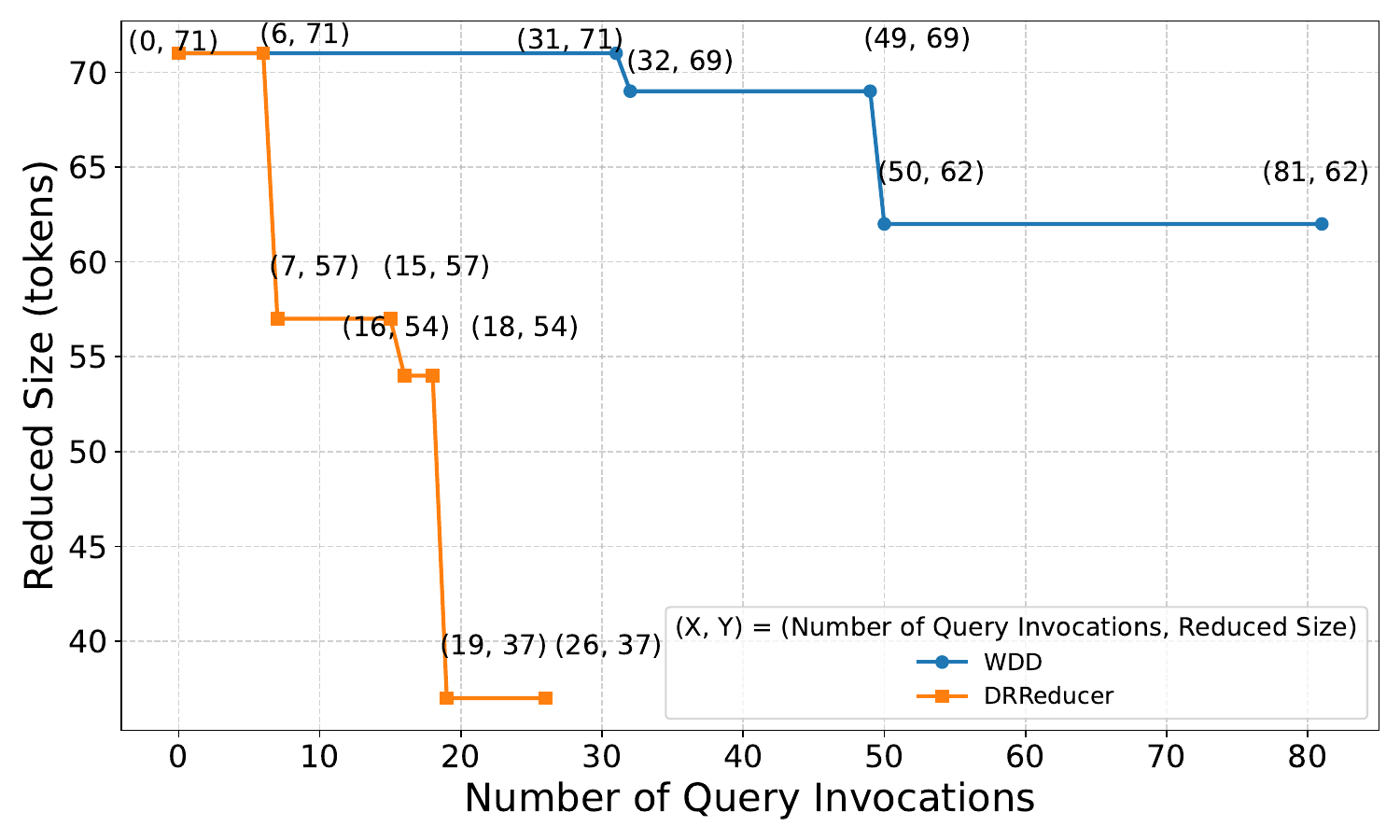}
    \caption{The overall reduction result of WDD and \toolName{}.}
    \label{fig:example-comparison}
\end{figure}



Program reduction takes as input a bug-triggering program $P$ and a property checker $\psi$ that determines whether a given program preserves the property of interest, and produces a smaller program $P'$ satisfying $\psi(P') = \psi(P)$. Figure~\ref{fig:mtv-hello-before} shows a Java program prior to reduction, where the property of interest is producing the ``Hello World'' output. Figure~\ref{fig:mtv-hello-after} presents the successfully reduced program produced by Weighted Delta Debugging (WDD)~\cite{zhou2024wdd}, which still prints ``Hello World''. Built on Perses~\cite{sun2018perses} - which leverages ANTLR grammars to transform programs into syntax trees — WDD is among the most advanced language-agnostic reduction techniques. It improves upon prior delta debugging approaches by assigning weights to syntactic elements (e.g., based on token count) and performing weight-aware partitioning, prioritizing the removal of smaller elements that are more likely to be irrelevant. Although WDD achieves state-of-the-art (SOTA) effectiveness and efficiency compared to other language-agnostic methods, it still exhibits two key limitations on this example.

\paragraph{Limitation 1: WDD generates invalid intermediate results, which limits efficiency.}

During node deletion, operations must often follow a strict order: removing elements out of sequence can violate syntactic or semantic dependencies. For example, in Figure~\ref{fig:mtv-hello-before}, the method \colorbox{softgreen}{\texttt{uselessFunc}} is invoked within \texttt{main}. Deleting its declaration before removing its call sites results in a compilation error. Thus, the statement \colorbox{softpink}{\texttt{uselessB = uselessFunc();}} must be eliminated before \colorbox{softgreen}{\texttt{uselessFunc}} can be safely removed.

To handle such ordering constraints, Perses employs a fixpoint iteration strategy: it repeatedly applies reduction passes until no further nodes can be deleted, ensuring that interdependent elements are eventually processed. In this example, Perses first attempts (and fails) to remove \colorbox{softgreen}{\texttt{uselessFunc}}, then successfully deletes \colorbox{softpink}{\texttt{uselessB = uselessFunc();}}, and continues iterating until reaching a 1-minimal state. WDD further improves this process by using weights to guide deletion order. However, as shown in Figure~\ref{fig:example-comparison}, even for this simple program, WDD requires 81 iterations, whereas our approach completes the reduction in only 26 iterations.

A closer inspection reveals that WDD frequently produces invalid intermediate states. In the first 31 iterations, all attempts fail the query script—for example, it removes \colorbox{softpurple!40}{\texttt{uselessA}} without eliminating its occurrences in \colorbox{softpurple!40}{\texttt{uselessA + uselessB}}. Only at iteration 32 does it successfully simplify the expression by removing \colorbox{softpurple!40}{\texttt{ + uselessB}}, reducing the token count to 69. Between iterations 33 and 49, despite two fixpoint cycles, no function-level elements are removed. The 50th iteration finally succeeds in deleting the statement \colorbox{softpink}{\texttt{int uselessB=uselessFunc();}}, after which further iterations again yield no progress.

The root cause is that WDD does not explicitly enforce the semantic ordering of deletions. \textit{Valid intermediate states must respect the topological order induced by dependency relationships among syntax nodes.} Ignoring this constraint leads to numerous failed attempts and significantly degrades efficiency, particularly for larger programs.

\begin{figure}[htpb]
\begin{subfigure}[t]{0.48\linewidth}
\centering
\includegraphics[width=\linewidth]{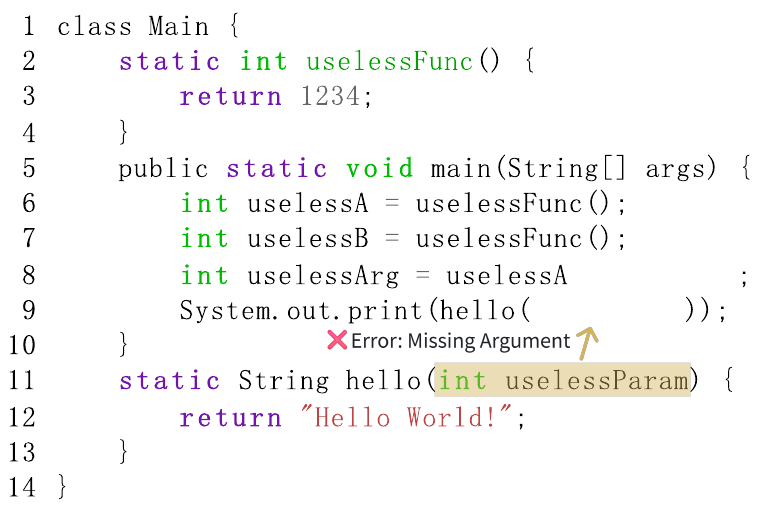}
\caption{Deleting an argument causes a compilation error (iteration 39.}
\label{fig:mtv-middle-wdd-1}
\end{subfigure}
\hfill
\begin{subfigure}[t]{0.48\linewidth}
\centering
\includegraphics[width=\linewidth]{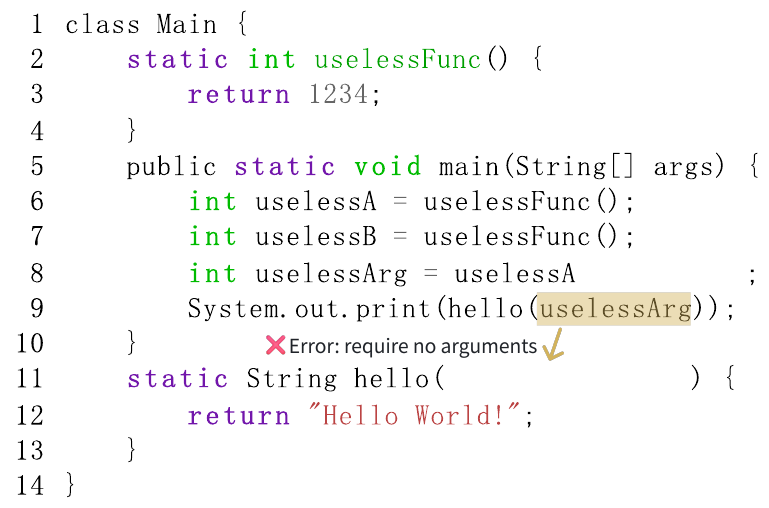}
\caption{Deleting a parameter causes a compilation error (iteration 42.}
\label{fig:mtv-middle-wdd-2}
\end{subfigure}
\caption{Two invalid intermediate results produced by WDD for Fig.~\ref{fig:mtv-hello-before}.}
\label{fig:mtv-middle-wdd}
\end{figure}

\begin{figure}[htbp]
    \centering
    \begin{subfigure}[t]{0.48\linewidth}
        \centering
        \includegraphics[width=\linewidth]{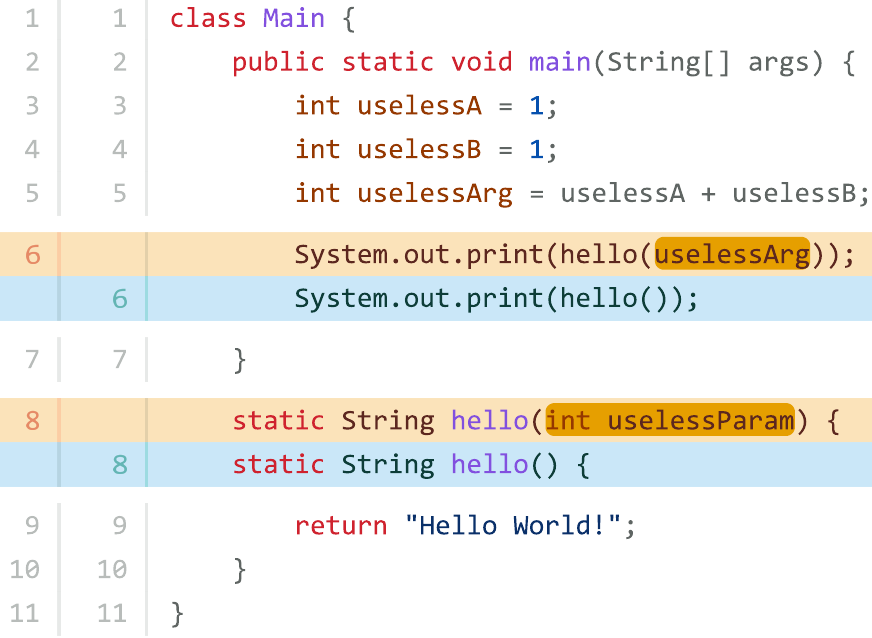}
        \caption{A valid intermediate program at iteration~16 (54 tokens), in which \toolName{} atomically removes a parameter and its corresponding argument together. }
        \label{fig:mtv-middle-16}
    \end{subfigure}
    \hfill
    \begin{subfigure}[t]{0.48\linewidth}
        \centering
        \includegraphics[width=\linewidth]{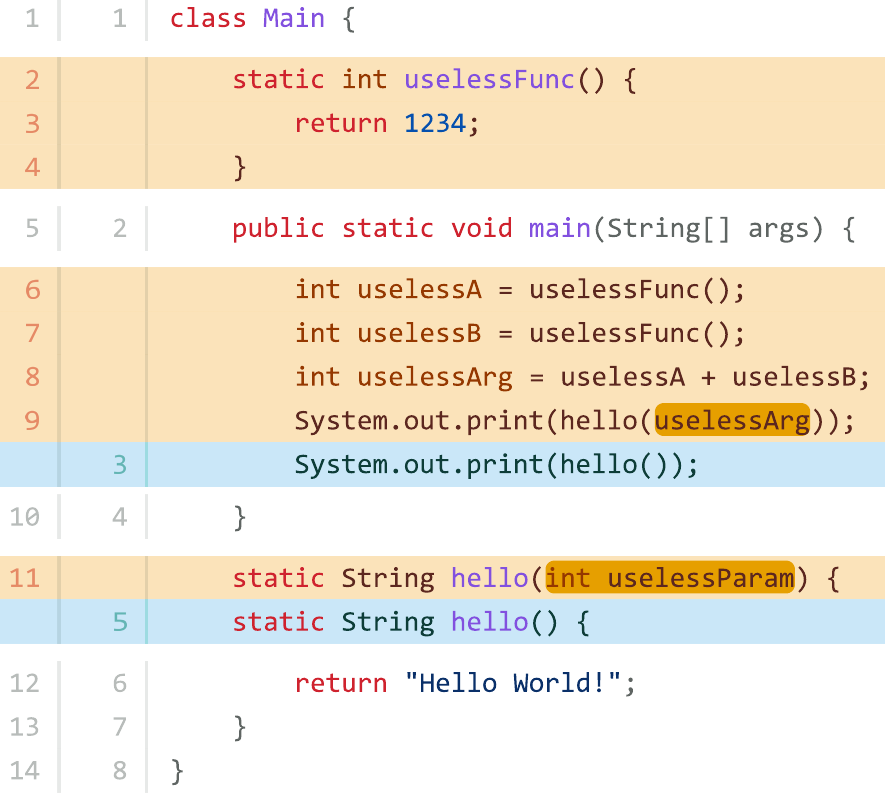}
        \caption{Final reduction result of Fig.~\ref{fig:mtv-hello-before} produced by \toolName{}. }
        \label{fig:mtv-after-drreduce}
    \end{subfigure}
    \caption{Intermediate and final reduction results produced by \toolName{} on the motivating example in Fig.~\ref{fig:mtv-hello-before}. Yellow shading indicates the lines that were present in the previous valid intermediate but have been deleted in this step.}
    \label{fig:mtv-drreduce-results}
\end{figure}

\paragraph{Limitation 2: WDD fails to handle mutually dependent elements, which limits effectiveness.}

Some syntax nodes form mutual semantic dependencies that prevent fixpoint reduction from making progress. We define a node that provides a dependency as a \textit{provider node} and one that consumes it as a \textit{user node}. While deleting a provider node typically breaks its user nodes, in some cases deleting the user node also causes errors.

Figure~\ref{fig:mtv-hello-before} illustrates such a scenario: \colorbox{softyellow}{\texttt{uselessParam}} and \colorbox{softyellow}{\texttt{uselessArg}} are mutually dependent, since function arguments and parameters must remain consistent for successful compilation. Removing either one isolately results in a compilation error, as illustrated in Figures~\ref{fig:mtv-middle-wdd-1} and~\ref{fig:mtv-middle-wdd-2}, which show two failed attempts by WDD at iterations~39 and~42. Consequently, nodes involved in such cycles cannot be deleted individually, and fixpoint iteration alone cannot break the impasse, leaving WDD stuck at a larger reduced size than is theoretically achievable.

\paragraph{Summary and our approach.}
The two limitations above share a common root cause: existing syntax-guided reducers treat syntax nodes as independent units and lack any awareness of the semantic dependencies between them. As a result, they generate invalid intermediate programs which fail to compile and also cannot remove elements involved in dependency cycles.

To address both limitations, we propose \textbf{\toolName{}: Augmenting Syntax-Guided Program Reduction with Dependency Reconstruction}. Given an input program, \toolName{} first identifies its semantic nodes (i.e., syntax nodes that participate in semantic dependencies, such as declarations, references, and parameters) and their dependencies.
It then iteratively deletes these semantic nodes using DDMin~\cite{ddmin}, and after each deletion, it applies \textit{dependency reconstruction} to proactively repair any broken references introduced by the removal. Dependency reconstruction takes two forms: when a deleted node is referenced by surviving users, \toolName{} substitutes a placeholder of the appropriate type at each reference site (the type remains visible and is sufficient to synthesize a placeholder in the program text even when the value does not); when a deleted node participates in a circular dependency, \toolName{} deletes the entire cycle. Concretely, when deleting \colorbox{softgreen}{\texttt{uselessFunc}}, its call sites on Lines~6 and~7 in Figure~\ref{fig:mtv-hello-before} are replaced with the constant expression \texttt{1}, as shown on Lines~3 and~4 in Figure~\ref{fig:mtv-middle-16}.
In another scenario, when deleting \colorbox{softyellow}{\texttt{uselessParam}}, the corresponding argument \colorbox{softyellow}{\texttt{uselessArg}} is deleted simultaneously to break the circular dependency. By keeping
intermediate programs compilable in both scenarios, \toolName{} substantially increases the rate at which the oracle accepts deletions, addressing both efficiency and effectiveness limitations of prior syntax-guided reducers.

As shown in Figure~\ref{fig:example-comparison}, \toolName{} reduces the example program from 71 to 37 tokens in 26 iterations, while WDD only reaches 62 tokens after 81 iterations; the final reduction result produced by \toolName{} is shown in Figure~\ref{fig:mtv-after-drreduce}. We describe the design of \toolName{} in detail in Section~\ref{sec:approach}.

%% file: approach.tex
\section{DRReduce}
\label{sec:approach}

\begin{figure}
    \centering
    \includegraphics[width=1\linewidth]{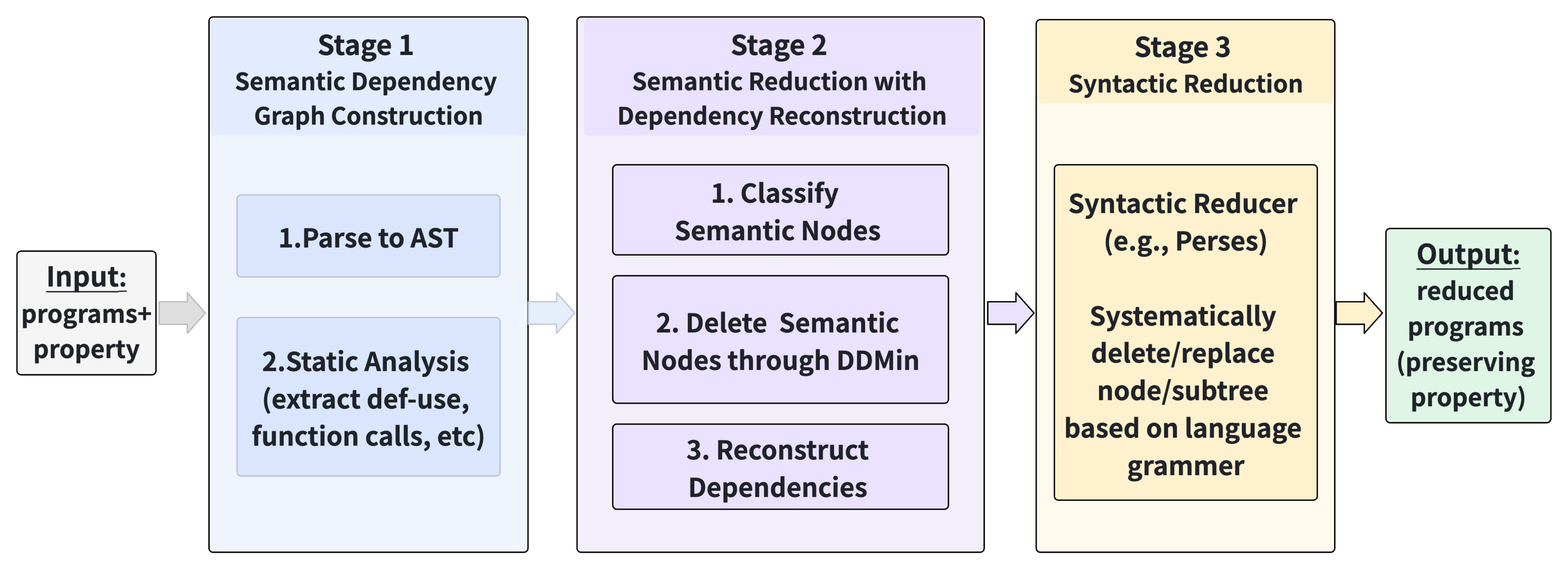}
    \caption{\toolName{}'s overall workflow}
    \label{fig:framework}
\end{figure}

The overall workflow of \toolName{} is shown in Figure~\ref{fig:framework}. Given a set of programs and a property of interest, \toolName{} proceeds in three stages.

In \textbf{Stage 1}, \toolName{} constructs a semantic dependency graph, a language-agnostic representation that can model dependencies across programs in different forms.
The graph construction strategy differs depending on whether semantic dependency information is explicitly available in the input. For intermediate representations (IRs) that already encode dependency information — such as IRs used for compiler testing~\cite{chaliasos2022finding, feng2025finding} or compilation IRs such as LLVM bitcode — \toolName{} constructs the graph directly without additional analysis. For source code or other program forms without explicit dependencies, \toolName{} first parses each program into an Abstract Syntax Tree (AST), then performs static analysis to extract semantic relationships such as definition-use chains, function call dependencies, and type constraints. In both cases, the resulting graph has the same structure and the subsequent reduction process is identical.

In \textbf{Stage 2}, semantic reduction with \emph{dependency reconstruction}, performs the actual semantic reduction and consists of three steps. (1) \toolName{} classifies semantic nodes that carry semantic significance.
(2) \toolName{} deletes semantic nodes using DDMin~\cite{ddmin}. (3) \toolName{} traverses the semantic dependency graph and reconstructs broken dependencies by rewiring them to valid replacements. By directly manipulating the program’s dependency structure, this stage produces a semantically valid reduced program that preserves the property of interest while remaining compilable and executable.


In \textbf{Stage 3}, \toolName{} passes the semantically reduced program to a syntactic reducer — Perses in our implementation — which performs further minimization by systematically attempting to delete or replace subtrees guided by the language grammar. Because Stage~2 has already produced a clean, semantically coherent core, Stage~3 encounters far fewer query invocations and converges faster than it would on the original input. The two stages are complementary, but the majority of \toolName{}'s effectiveness gain over prior reducers comes from Stage~2's \emph{dependency reconstruction}--- the core contribution of this work --- which performs reductions that no purely syntactic search can achieve, as illustrated in our motivating example (Section~\ref{sec:motivation}) and further confirmed in our ablation study (Section~\ref{subsec:rq3}).

Algorithm~\ref{alg:sdrdr} presents the full procedure of Stage 2: semantic reduction with \textit{dependency reconstruction}.
Given a set of programs $\mathcal{P}$, a property checker $\psi$, and the semantic dependency graph generated in Stage~1, this procedure proceeds as a single loop driven by the DDMin search~\cite{ddmin},
via the routines \textsc{InitDDMin}, \textsc{NextCandidate}, and \textsc{UpdateDDMin}. The search maintains internal state that determines the next deletion candidate $\mathcal{RC}_{del}$ based on the accepted and rejected attempts, rather than enumerating subsets in a fixed order. Each iteration (Lines~7--13) applies one such candidate to a working copy of the program, performs dependency reconstruction, and queries the oracle $\psi$.

When a deletion is accepted, all nodes in $\mathcal{RC}_{del}$ are removed from $G$, along
with every edge incident to them. Second, the edges that previously connected those nodes to remaining user nodes are rewired to the placeholder expressions introduced by reconstruction. After this update, semantic-node classification is re-run on the modified graph to provide any new reduction opportunities exposed by the deletion. \toolName{} continues until the ddmin search exhausts its candidates.



\begin{algorithm}
\caption{Semantic Reduction with Dependency Reconstruction}
\label{alg:sdrdr}
\begin{algorithmic}[1]
\Require $G$, Set of programs $\mathcal{P} = \{P_1, P_2, \ldots, P_n\}$, property $\psi$
\Ensure Reduced programs $\mathcal{P}' = \{P_1', P_2', \ldots, P_n'\}$ each satisfying $\psi$
\State $\mathcal{RC}_{sem} \gets \textsc{ClassifySemanticNodes}(G)$
\State $\mathcal{P}' \gets \mathcal{P}$
\State $\textit{state} \gets \textsc{InitDDMin}(\mathcal{RC}_{sem})$
\While{$\textit{state}$ has unexplored candidates}
    \State $\mathcal{RC}_{del} \gets \textsc{NextCandidate}(\textit{state})$
    \State $\mathcal{P}_{test} \gets \text{Copy}(\mathcal{P'})$
    \State Delete all nodes in $\mathcal{RC}_{del}$ from $P_{test}$
    \State $\textsc{ReconstructDependencies}(\mathcal{P}_{test}, G, \mathcal{RC}_{del})$
    \If{$P_{test} \models \psi$}
        \State $\mathcal{P'} \gets \mathcal{P}_{test}$
        \State $G \gets \textsc{UpdateGraph}(G, \mathcal{RC}_{del})$ \Comment{Remove deleted nodes; rewire edges to reconstructed
                 placeholders}
        \State $\mathcal{RC}_{sem} \gets \textsc{ClassifySemanticNodes}(G)$
        \State $\textit{state} \gets \textsc{UpdateDDMin}(\textit{state},
                \mathcal{RC}_{sem}, \text{accept})$
    \Else
        \State $\textit{state} \gets \textsc{UpdateDDMin}(\textit{state},
                \mathcal{RC}_{sem}, \text{reject})$
    \EndIf
\EndWhile


\State \Return $\mathcal{P}'$
\end{algorithmic}
\end{algorithm}

The remainder of this section describes the two core steps of Algorithm~\ref{alg:sdrdr} in detail: classifying semantic nodes (Section~\ref{sec:approach-select}) and reconstructing dependencies (Section~\ref{sec:approach-reconstruct}).

\subsection{Classifying Semantic Nodes}
\label{sec:approach-select}

Effective dependency reconstruction requires identifying which syntax nodes participate in semantic relationships in the first place. Not all syntax nodes are equally relevant to semantics. To focus reduction on semantically meaningful nodes, \toolName{} classifies nodes into three categories based on their semantic role:

\begin{itemize}
\item \emph{Provider}: the node defines a semantic entity that other nodes depend on (e.g., a function declaration, a variable declaration, a parameter declaration).
\item \emph{User}: the node references a semantic entity defined by other nodes (e.g., a function call, a variable reference, an argument at a call site).
\item \emph{Conditioner}: the node neither defines nor references a semantic entity directly, but conditions whether other dependencies remain valid (e.g., a \texttt{public} modifier, whose removal can invalidate references to the enclosing entity from other packages).
\end{itemize}

This classification extends classical def-use analysis and generalizes its notion. Providers and users naturally capture def-use relationships, but they also capture other forms of semantic dependency that def-use analysis does not, such as the positional correspondence between formal parameters and their actual arguments at call sites, even though no variable is being defined or referenced.
 The \emph{conditioner} category has no def-use relations and captures syntactic structures whose removal would invalidate many provider-user relationships. Consider the nodes of a Java method declaration: annotation nodes, modifier nodes, and parameter nodes. Annotation nodes reference a class declaration and therefore act purely as \emph{users}. Modifier nodes (e.g., \texttt{public}, \texttt{static}) determine whether the method can be referenced externally, making them \emph{conditioners}. Parameter nodes are both \emph{users} and \emph{providers}: they have bidirectional dependencies with argument expressions at call sites, and also define entities consumed by nodes within the method body.

\toolName{} models semantic relations between Provider and User nodes as directed edges $(u, p) \in E$, where $u \in N_{sem}$ is a \emph{user} that directly requires $p \in N_{sem}$, a \emph{provider}, for its semantic validity. The resulting dependency graph $G = (N_{sem}, E)$ captures all such edges over the selected semantic nodes. The deletion behavior of nodes follows naturally: removing $p$ renders $u$ semantically invalid, while removing $u$ eliminates the dependency on $p$, making $p$ a candidate for removal if no other \emph{user} depends on it. For example, the method \colorbox{softgreen}{\texttt{hello}} directly requires the parameter \colorbox{softyellow}{\texttt{int uselessParam}} it corresponds to, yielding the edge (\colorbox{softgreen}{\texttt{hello}} ,\ \colorbox{softyellow}{\texttt{int uselessParam}}).  In C, a call site is a \emph{user} of the function declaration it references, and a function's forward declaration is a \emph{user} of its definition.

\toolName{} selects only \emph{provider} and \emph{user} nodes as reduction candidates, since their semantic roles are well-defined and can be precisely captured by semantic dependencies. \emph{conditioner} nodes, whose semantic effects are more complex and context-dependent, are delegated to the syntax-guided reducer in Stage 3. For instance, in C, a function's return type node is treated as a \emph{conditioner} and excluded from \toolName{}'s reduction candidates; though it remains a candidate for the syntax-guided reducer.

\subsection{Reconstructing Dependencies}
\label{sec:approach-reconstruct}
When a \emph{provider} node in a reduction candidate is deleted atomically through DDMin, any \emph{user} nodes that depend on the deleted provide node will produce compilation errors. To prevent this, \methodName{} applies \textit{dependency reconstruction} before committing any deletion: for each provider $p$ scheduled for deletion, every user $u$ such that $(u, p) \in E$ that has not itself been deleted is modified to remove its dependency on $p$, keeping the intermediate program compilable.

Dependency reconstruction takes two forms.
\begin{itemize}
\item \textit{Default value replacement} substitutes the \emph{user} node with the default value of its expected type. In Figure~\ref{fig:mtv-hello-before}, after scheduling \texttt{uselessFunc} for deletion, \methodName{} reconstructs the dependency at Lines~6 and~7 by replacing \texttt{uselessFunc} with \texttt{1}. The modified intermediate program compiles successfully, allowing the deletion to proceed without error, as shown in Figure~\ref{fig:mtv-middle-16}. This \textit{Default value replacement} operation modifies program semantics but preserves their semantics to the extent possible.
\item \textit{Associated semantic structure deletion} removes groups of nodes that share a mutual semantic relationship and cannot be deleted independently without breaking compilation. One example is the relationship between a formal parameter in a function declaration and its corresponding actual argument at every call site: deleting one without the other produces a signature mismatch. As illustrated in Figure~\ref{fig:mtv-after-drreduce}, this \textit{Associated semantic structure deletion} operation enables  \toolName{} to remove both \colorbox{softyellow}{\texttt{uselessParam}} and \colorbox{softyellow}{\texttt{uselessArg}} together.


\end{itemize}

The choice between the two forms is determined by the dependency structure: \toolName{} applies default-value replacement when the deleted node has surviving users that reference its value, and associated structure deletion when the deleted node participates in a circular dependency where no individual node can be removed without breaking another, such as parameter-argument pairs.


\subsubsection*{Semantic validity guarantees.}
Dependency reconstruction guarantees the \emph{compilability} of every intermediate program but not \emph{behavioral equivalence} to the original. By substituting type-compatible default values for broken references or deleting cyclic dependencies, the modified program may compute different values than the original, but it always compiles.

This weaker guarantee does not compromise the soundness of \toolName{}: every intermediate is independently checked against the property $\psi$, and modifications that violate $\psi$ are rejected exactly as any other failed reduction would be. The role of reconstruction is to ensure that $\psi$ can be \emph{evaluated} --- the program compiles --- not to ensure that it is \emph{satisfied}. This decouples compilability from property preservation, which syntax-guided reducers conflate: prior tools treat compilation failures as evidence of failed reduction even though such failures reveal nothing about $\psi$.

In practice, most bug-triggering properties (compiler crashes, miscompilations of structural patterns, type-checker false positives and false negatives) depend on syntactic or type-level features rather than concrete runtime values, so default-value reconstruction preserves $\psi$ in the vast majority of cases. When it does not, the oracle rejects the modification and \toolName{} backtracks. Our evaluation (Section~\ref{subsec:rq1}) identifies one such case (\cf{691}), where the bug-triggering property is sensitive to a specific type annotation that default-value substitution destroys; this is the only program in our 28-program benchmark on which \toolName{} underperforms a syntax-guided baseline.

\subsection{Language-Specific Implementation}
\label{sec:impl}

While Algorithm~\ref{alg:sdrdr} describes a general semantics-guided reduction process, its full realization requires semantic analysis and dependency reconstruction capabilities that vary across languages and program forms. \toolName{} currently supports two targets: C and Java source code. We choose these two languages for two reasons. First, both are widely used in compiler testing and have well-established benchmarks and baselines for program reduction, enabling direct comparison with prior work. Second, JetBrains' Program Structure Interface (PSI)~\cite{psi} provides unified APIs for dependency analysis, type resolution, and refactoring operations across both languages, which we leverage as the foundational infrastructure for semantic analysis and transformation in \toolName{}. Table~\ref{tbl:semantic-nodes} summarizes the semantic nodes for each language, and Table~\ref{tbl:reconstruction} details the dependency reconstruction strategies.



\paragraph{Extending \toolName{} to other languages.} Adding support for a new language requires three pieces of language-specific information: (1) a parser and AST representation, (2) a mapping from language constructs to the provider, user, and conditioner roles in Table~\ref{tbl:semantic-nodes}, and (3) default values for each primitive type. The mapping is small (our C and Java implementations each define fewer than ten node categories) and follows directly from the language's type system; default values are trivially available in any typed language. Notably, no language-specific transformation rules are required, since the reduction logic itself remains unchanged across languages. Languages already integrated with PSI (e.g., Kotlin, Python, Rust, Go) can reuse the existing front-end with no new analysis infrastructure.

\input{tables/approach/impl.tex}
\begin{table}[htbp]
    \centering
    \caption{Semantic node categories identified by \toolName{} for C and Java.}
    \label{tbl:semantic-nodes}
    \begin{tabular}{p{2cm} p{11cm}}
        \toprule
         & \textbf{Semantic Nodes}\\
        \midrule
        C &  Files, Declarations, Statements,
            Function declaration and its definition,
            parameter declaration and its corresponding arguments \\
        \midrule
        Java & Files, Statements, Named Nodes, Parameters,
         Parameter declaration and its corresponding arguments \\
        \bottomrule
    \end{tabular}
\end{table}

\begin{table}[htbp]
    \centering
    \caption{Dependency Reconstruction Strategies for C and Java}
    \label{tbl:reconstruction}
    \begin{threeparttable}
    \begin{tabular}{p{1.5cm} p{2.5cm} p{2.5cm} p{5cm}}
        \toprule
        \textbf{Target} & \textbf{Provider} & \textbf{User} & \textbf{Reconstruction Strategy} \\
        \midrule
        \multirow{6}{*}{C}
            & Type / Struct             & Type reference   & Replace with \texttt{void***} \\ \cline{2-4}
            & \multirow{2}{*}{Function} & Expression       & Replace with \texttt{(func\_type) 0} \\
            &                           & Call expression  & Replace with \texttt{(return\_type) default}\tnote{a} \\ \cline{2-4}
            & Variable / Parameter      & Expression       & Replace with \texttt{default}\tnote{a} \\ \cline{2-4}
            & Goto label                & Goto statement   & Delete User node \\ \cline{2-4}
            & Parameter / Argument & Argument / Parameter & Delete User node \\
        \midrule
        \multirow{5}{*}{Java}
            & Class                     & Type reference   & Replace with \texttt{Object} \\ \cline{2-4}
            & \multirow{2}{*}{Method}   & Expression       & Replace with \texttt{(func\_type) null}\tnote{b} \\
            &                           & Call expression  & Replace with \texttt{(return\_type) default}\tnote{c} \\ \cline{2-4}
            & Variable / Parameter      & Expression       & Replace with \texttt{(expected\_type) default}\tnote{c} \\ \cline{2-4}
            & Parameter / Argument & Argument / Parameter & Delete User node \\
        \bottomrule
    \end{tabular}
    \begin{tablenotes}
        \item[a] For C, the default value is determined by the expected or return type:
        \texttt{1} for integer types, \texttt{0} for pointer types.
        \item[b] Java does not have first-class function types; method references are
        typically typed as functional interface types.
        \item[c] For Java, the default value is \texttt{0} for primitive types and
        \texttt{null} for reference types.
    \end{tablenotes}
    \end{threeparttable}
\end{table}

\subsubsection{C Source Code.}
\label{sec:impl-c}

For C programs, \toolName{} performs semantic analysis using JetBrains' Program Structure Interface (PSI)~\cite{psi}, a language-aware API provided by IntelliJ CLion that represents source code as a structured tree of typed nodes, each carrying syntactic and semantic information such as type bindings, reference resolution, and scope. \toolName{} leverages PSI to identify semantic dependencies between nodes without requiring a custom parser or type checker.

The selected semantic nodes are Files, Declarations, and Statements. Declarations cover struct declarations, function declarations, variable declarations, struct members, and parameter declarations. Expression nodes are excluded because dependency reconstruction for expressions requires non-trivial type inference, which we leave for future work. Two kinds of node pairs are grouped together: (1) a function's forward declaration and its definition --- since deleting either alone leaves an unresolved symbol --- and (2) a parameter declaration and its corresponding argument expressions at all call sites --- since function signatures and call sites must remain consistent. The reconstruction strategies are summarized in Table~\ref{tbl:reconstruction}. For example, if a deleted function \texttt{foo} returns \texttt{int}, then a remaining call \texttt{foo()} is replaced with \texttt{(int) 1}, according to the reconstruction rule between Function and Call expressions in Table~\ref{tbl:reconstruction}.

\subsubsection{Java Source Code.}
\label{sec:impl-java}

For Java programs, \toolName{} performs semantic analysis using PSI provided by JetBrains' IntelliJ IDEA~\cite{psi}, which offers Java-specific capabilities such as type hierarchy resolution, method override detection, and reference tracking across class boundaries.

The selected semantic nodes are Files, Statements, Named Nodes, and Parameters. Named Nodes are elements referenceable by name, including classes, methods, and fields. Unlike C, Java does not require forward declarations, so no declaration-definition pairs exist. Parameter declarations and their corresponding argument expressions at all call sites are grouped together, since method signatures and call sites must remain consistent.

Java's inheritance structure introduces additional semantic complexity. When a superclass is deleted, its subclasses must be updated to inherit from the grandparent, and overriding methods that reference the deleted superclass must be removed or redirected. Correctly propagating such changes across the full class hierarchy at the source-code level is beyond our current capabilities and is left for future work. The reconstruction strategies currently supported are summarized in Table~\ref{tbl:reconstruction}. For example, suppose a class \texttt{MyClass} is deleted while the program still contains a method that uses \texttt{MyClass} as a parameter type, such as \texttt{void foo(MyClass arg) \{ ... \}}. According to the rules in Table~\ref{tbl:reconstruction}, \texttt{MyClass} is replaced with its supertype \texttt{Object}, yielding the reconstructed method signature \texttt{void foo(Object arg) \{ ... \}}.


%% file: experiment-setup.tex
\section{Experiment Setup}
\label{sec:experiment}

\subsection{Dataset}
\newcommand{\cfName}{Checker Framework}
\newcommand{\naName}{Null Away}
\newcommand{\ecjName}{Eclipse Java Compiler}

Table~\ref{tbl:all-data-dresult} summarizes the dataset used to evaluate \toolName{}, covering bug-triggering programs in C and Java languages across multiple compilers and data sources.

For C, the dataset contains 16 bug-triggering programs sourced from the Perses dataset~\cite{sun2018perses}, covering 10 GCC bugs and 6 Clang bugs. For Java, the dataset contains 12 programs covering 12 bugs across three compilers: 4 programs target \cfName{} bugs from Specimin~\cite{specimin}; 2 programs target \ecjName{} and 2 target JDK bugs, both from Perses; and an additional 4 programs target JDK bugs from newly collected data. Note that \cfName{} is a type-checking plugin for Java compilers.

Several programs were excluded from the original datasets for the following reasons, as detailed in the footnotes of Table~\ref{tbl:all-data-dresult}. First, manual inspection revealed that for 3 Clang bugs and 1 GCC bug, the reduced programs from the original reports trigger a different bug than the one reported. Second, 1 \ecjName{} bug was excluded because the compiler version required to reproduce the bug is no longer publicly archived. Third, 1 \naName{} bug was excluded because it is a false negative tied to a specific location; a user query script cannot be accurately written in this case, and automatic reduction is not meaningful since such location-specific false negatives can already be reduced manually.

\begin{table}[!ht]
    \centering
    \caption{The bug-triggering program dataset used to evaluate \toolName{}.}
    \label{tbl:all-data-dresult}
    \begin{threeparttable}
    \begin{tabular}{l l r r p{3.5cm}}
        \toprule
        \textbf{Program Type} & \textbf{Compiler} & \textbf{\#Programs} &
        \textbf{\#Bugs} & \textbf{Data Source} \\
        \midrule
        \multirow{2}{*}{C}
            & GCC   & 10 & 10 & \multirow{2}{*}{Perses~\cite{sun2018perses}} \\
            & Clang & 6  & 6  & \\
        \midrule
        \multirow{4}{*}{Java}
            & \cfName{}  & 4 & 4 & Specimin~\cite{specimin} \\
            \cmidrule(lr){2-5}
            & \ecjName{} & 2 & 2 & \multirow{2}{*}{Perses~\cite{sun2018perses}} \\
            & JDK        & 2 & 2 & \\
            \cmidrule(lr){2-5}
            & JDK        & 4 & 4 & Newly Collected \\
        \bottomrule
    \end{tabular}
    \begin{tablenotes}
        \small
        \item[$\dagger$] 3 Clang bugs and 1 GCC bug from the Perses dataset were removed upon manual inspection, which revealed that the reduced programs from the original reports trigger a different bug than the one reported.
        \item[$\ddagger$] 1 \ecjName{} bug from the Perses dataset was removed because the compiler version required to reproduce the bug is no longer publicly archived.
        \item[$\S$] 1 \naName{} bug from the Specimin dataset was removed because it is a location-sensitive false negative for which no general-purpose query script can be written, making automated reduction ill-defined.
    \end{tablenotes}
    \end{threeparttable}
\end{table}

\subsection{Experiment Configuration}

All experiments were conducted on a machine running Ubuntu~24.04.3~LTS (64-bit) with Linux kernel~6.14.0-37-generic, powered by an AMD Ryzen~9 9950X 16-core processor (32~threads) and 64~GiB of RAM. Each bug-triggering program was reduced using a single thread to ensure fair and reproducible comparisons across all tools.

In our implementation of \toolName{}, we adopt Perses~\cite{sun2018perses} as the syntactic reducer in Stage~3. We make this choice because Perses is the foundational syntax-guided reducer upon which the other syntax-guided baselines in our evaluation (WDD~\cite{zhou2024wdd} and CDD~\cite{perses-cdd}) are built. Using the same backend across all tools allows us to fairly compare the impact of \toolName{}'s dependency reconstruction against the improvements that WDD and CDD make on top of Perses~\cite{sun2018perses}.

\subsection{Research Questions}
Using the above dataset, we evaluate whether \toolName{}'s effectiveness and efficiency compare with those of state-of-the-art (SOTA) program reduction approaches. Accordingly, we formulate the following three Research Questions (RQs):

\textbf{RQ1. How does \toolName{} compare to state-of-the-art syntax-guided reduction tools in terms of reduction effectiveness and efficiency?} Specifically, does \toolName{} produce smaller reduced programs and require less reduction time than existing tools? Existing syntax-guided reduction tools treat syntax nodes as independent units, ignoring the semantic dependencies between them. This design choice leads to two main issues: (1) tools fail to reduce elements with mutual semantic dependencies, limiting how small the output can become; and (2) without respecting dependency ordering, tools frequently generate invalid intermediate programs, triggering unnecessary query invocations and increasing reduction time. This RQ evaluates whether \toolName{} can address these shortcomings, producing smaller outputs in less time on C and Java bug-triggering programs.

\textbf{RQ2. How does \toolName{} compare to state-of-the-art reducers that incorporate language-specific transformations in terms of reduction effectiveness and efficiency?} Tools that incorporate language-specific transformations, such as CReduce and Latra, achieve exceptionally compact outputs by applying extensive, hand-crafted, semantics-aware transformations tailored to a specific language.
In contrast, \toolName{} relies solely on a language-agnostic strategy of dependency reconstruction, with no language-specific rules. This RQ explores whether \toolName{}'s general approach incurs a measurable cost: does its simpler, uniform design yield comparable reduction quality or speed to that of highly engineered, special-purpose tools?

\textbf{RQ3 (Ablation). Is dependency reconstruction a necessary component of \toolName{}, and what is its overall impact on effectiveness and efficiency?} Dependency reconstruction ensures the compilability of intermediate results by substituting references to deleted nodes with default values. While this prevents unnecessary query invocations due to compilation failures, it also introduces additional code changes that may affect both the final program size and the time required to reach a fixpoint. This RQ isolates the contribution of dependency reconstruction by measuring its impact on both effectiveness and efficiency, and assesses whether its benefits consistently outweigh its overhead.

%% file: results.tex
\section{Results}
\label{sec:results}
\input{rq1}
\input{rq2}
\input{rq3}

%% file: rq1.tex
\subsection{RQ1: Comparison between \toolName{} and SOTA  syntax-guided reduction tools}
\label{subsec:rq1}

\newcommand{\mybest}[1]{\cellcolor{green!40}{#1}}
\newcommand{\mybestt}[1]{\cellcolor{blue!20}{#1}}

\begin{table*}[htbp]
\centering
\caption{Reduction results on C and Java bug-triggering programs compared with syntax-guided reduction
tools. Q = number of query invocations, T = time (seconds), R = reduced size (tokens). $C = \frac{R_{\text{\toolName{}}} - R_{\text{Baseline}}}{R_{\text{Baseline}}} \times 100\%$ (negative values indicate \toolName{} produces smaller outputs than
the baseline).
Best reduced size per bug is highlighted in \colorbox{green!40}{green}.
Best time per bug is highlighted in \colorbox{blue!20}{blue}.
}
\label{tbl:rq1}
\resizebox{\textwidth}{!}{%
\begin{tabular}{l r | rrr | rrr | rrr | rrr | rrr}
\toprule
\multirow{2}{*}{\textbf{Bug ID}} &
\multirow{2}{*}{\textbf{Original}} &
\multicolumn{3}{c|}{\textbf{\toolName{}\textsubscript{Perses}}} &
\multicolumn{3}{c|}{\textbf{Perses~\cite{sun2018perses}}} &
\multicolumn{3}{c|}{\textbf{WDD~\cite{zhou2024wdd}}} &
\multicolumn{3}{c|}{\textbf{CDD~\cite{perses-cdd}}} &
\multicolumn{3}{c}{\textbf{C(\%) w.r.t.}} \\
\cmidrule(lr){3-5} \cmidrule(lr){6-8} \cmidrule(lr){9-11}
\cmidrule(lr){12-14} \cmidrule(lr){15-17}
              &       & Q     & T              & R            & Q    & T              & R   & Q    & T              & R            & Q    & T    & R            & Perses    & WDD       & CDD       \\\midrule
\clang{22382} & 21068 & 3459  & 118            & \mybest{95}  & 1445 & \mybestt{91}   & 256 & 1991 & 130            & 113          & 1998 & 135  & 113          & -62.9\%   & -15.9\%   & -15.9\%   \\ \hline
\clang{22704} & 93032 & 2888  & 223            & \mybest{63}  & 1413 & \mybestt{204}  & 215 & 1904 & 324            & 99           & 1866 & 316  & 97           & -70.7\%   & -36.4\%   & -35.1\%   \\ \hline
\clang{23353} & 30196 & 3763  & 236            & \mybest{72}  & 1881 & \mybestt{192}  & 283 & 2569 & 327            & \mybest{72}  & 2639 & 330  & \mybest{72}  & -74.6\%   & 0.0\%     & 0.0\%     \\ \hline
\clang{25900} & 78960 & 3266  & 279            & \mybest{107} & 1851 & \mybestt{240}  & 318 & 2659 & 311            & 148          & 2702 & 325  & 146          & -66.4\%   & -27.7\%   & -26.7\%   \\ \hline
\clang{26760} & 32958 & 2612  & \mybestt{151}            & \mybest{50}  & 1805 & 152  & 239 & 2161 & 176            & 64           & 2227 & 187  & 94           & -79.1\%   & -21.9\%   & -46.8\%   \\ \hline
\clang{27747} & 19915 & 2344  & \mybestt{108}  & \mybest{115} & 1430 & 114            & 269 & 2214 & 152            & 141          & 2211 & 152  & 141          & -57.2\%   & -18.4\%   & -18.4\%   \\ \hline\midrule
\gcc{59903}   & 57581 & 7127  & \mybestt{540}            & \mybest{328} & 3139 & 747  & 609 & 5422 & 817            & 394          & 5208 & 832  & 394          & -46.1\%   & -16.8\%   & -16.8\%   \\ \hline
\gcc{60116}   & 75224 & 10944 & 1765           & 433          & 3572 & \mybestt{469}  & 793 & 5062 & 651            & \mybest{393} & 5040 & 671  & \mybest{393} & -45.4\%   & 10.2\%    & 10.2\%    \\ \hline
\gcc{61383}   & 32449 & 7777  & 2503           & \mybest{258} & 2728 & \mybestt{2172} & 508 & 4185 & 3293           & 303          & 5032 & 3361 & 323          & -49.2\%   & -14.9\%   & -20.1\%   \\ \hline
\gcc{61917}   & 85359 & 8307  & 1698           & \mybest{127} & 2720 & \mybestt{291}  & 454 & 3841 & 576            & 147          & 3848 & 507  & 147          & -72.0\%   & -13.6\%   & -13.6\%   \\ \hline
\gcc{64990}   & 16623 & 4530  & 2042           & \mybest{163} & 2683 & \mybestt{1808} & 519 & 3503 & 2022           & 244          & 3516 & 2119 & 181          & -68.6\%   & -33.2\%   & -9.9\%    \\ \hline
\gcc{65383}   & 43942 & 12945 & \mybestt{3098}           & \mybest{122} & 2045 & 4112 & 384 & 2721 & 5232           & 144          & 2663 & 4960 & 156          & -68.2\%   & -15.3\%   & -21.8\%   \\ \hline
\gcc{66186}   & 47481 & 5782  & 2347           & 236          & 2253 & \mybestt{1090} & 488 & 3097 & 1422           & 310          & 2722 & 4999 & \mybest{153} & -51.6\%   & -23.9\%   & 54.2\%    \\ \hline
\gcc{70127}   & 12358 & 4228  & 4186           & \mybest{107} & 1817 & \mybestt{525}  & 370 & 2715 & 941            & 311          & 2757 & 956  & 311          & -71.1\%   & -65.6\%   & -65.6\%   \\ \hline
\gcc{70586}   & 14626 & 10675 & 6145           & \mybest{346} & 3568 & \mybestt{1768} & 904 & 6014 & 3008           & 402          & 6422 & 3390 & 380          & -61.7\%   & -13.9\%   & -8.9\%    \\ \hline
\gcc{71626}   & 6133  & 262   & 12             & \mybest{46}  & 120  & \mybestt{8}    & 49  & 249  & 16             & \mybest{46}  & 721  & 35   & \mybest{46}  & -6.1\%    & 0.0\%     & 0.0\%     \\ \hline\midrule
\cf{577}      & 382   & 782   & \mybestt{324}           & \mybest{209} & 548  & 351            & 306 & 532  & 347  & 306          & 536  & 356  & 306          & -31.7\%   & -31.7\%   & -31.7\%   \\ \hline
\cf{689}      & 2328  & 1863  & 408            & \mybest{104} & 231  & \mybestt{174}  & 209 & 393  & 292            & 146          & 386  & 290  & 146          & -50.2\%   & -28.8\%   & -28.8\%   \\ \hline
\cf{691}      & 25477 & 2458  & \mybestt{936}  & 454          & 5782 & 4736           & 601 & 7700 & 4773           & \mybest{95}  & 6535 & 4131 & 601          & -24.5\%   & 377.9\%   & -24.5\%   \\ \hline
\cf{4614}     & 394   & 118   & \mybestt{113}  & \mybest{67}  & 258  & 236            & 74  & 217  & 188            & 74           & 235  & 207  & 74           & -9.5\%    & -9.5\%    & -9.5\%    \\ \hline
\ecj{352665}  & 1142  & 112   & \mybestt{48}   & \mybest{44}  & 295  & 126            & 143 & 457  & 199            & 102          & 444  & 192  & 102          & -69.2\%   & -56.9\%   & -56.9\%   \\ \hline
\ecj{404146}  & 348   & 193   & \mybestt{64}   & \mybest{75}  & 435  & 165            & 155 & 379  & 145            & 155          & 382  & 145  & 155          & -51.6\%   & -51.6\%   & -51.6\%   \\ \hline
\jdk{8068399} & 447   & 246   & 25   & \mybest{63}  & 208  & 27             & 69  & 229  & \mybestt{23}             & 69           & 228  & \mybestt{23}   & 69           & -8.7\%    & -8.7\%    & -8.7\%    \\ \hline
\jdk{8145466} & 462   & 111   & \mybestt{14}   & \mybest{40}  & 197  & 29             & 49  & 173  & 22             & 49           & 172  & 22   & 49           & -18.4\%   & -18.4\%   & -18.4\%   \\ \hline
\jdk{8271954} & 1617  & 1253  & 1218           & \mybest{199} & 736  & 653  & 217 & 741  & \mybestt{512}            & 235          & 770  & 533  & 235          & -8.3\%    & -15.3\%   & -15.3\%   \\ \hline
\jdk{8272562} & 1537  & 1060  & 4741           & \mybest{124} & 522  & \mybestt{2357} & 134 & 513  & 2688           & 140          & 498  & 2840 & 140          & -7.5\%    & -11.4\%   & -11.4\%   \\ \hline
\jdk{8293941} & 1837  & 541   & 1453           & \mybest{123} & 363  & \mybestt{642}  & 144 & 486  & 3078           & 167          & 382  & 694  & 152          & -14.6\%   & -26.3\%   & -19.1\%   \\ \hline
\jdk{8331717} & 1497  & 1325  & 5755           & \mybest{136} & 855  & 2776 & 212 & 871  & \mybestt{1801}           & 212          & 874  & 1840 & 212          & -35.8\%   & -35.8\%   & -35.8\%   \\ \hline\midrule
Median        & 15625 & 2535  & 474            & \mybest{119} & 1438 & \mybestt{321}  & 263 & 2076 & 430            & 147          & 2105 & 432  & 150          & -54.8\%   & -19.0\%   & -20.7\%   \\ \hline
Mean          & 25192 & 3606  & 1448           & \mybest{154} & 1604 & \mybestt{938}  & 320 & 2250 & 1195           & 181          & 2251 & 1234 & 192          & -51.9\%   & -14.9\%   & -19.8\%   \\ \hline

\bottomrule
\end{tabular}%
}
\end{table*}

We compare \toolName{} against three state-of-the-art syntax-guided reducers: Perses~\cite{sun2018perses}, WDD~\cite{zhou2024wdd}, and CDD~\cite{perses-cdd}. These baselines span two generations of syntax-guided reduction: Perses (2018) establishes the grammar-based hierarchical reduction paradigm and remains the most widely adopted baseline, while WDD and CDD (2025) represent the latest advances, each improving upon the underlying delta debugging algorithms that Perses relies on. We adopt WProbDD --- WDD's stronger variant according to the original paper --- in our evaluation. Detailed descriptions of each baseline are provided in Section~\ref{sec:related} (Related Work).

In our implementation of \toolName{}, we adopt Perses as the syntactic reducer in Stage~3, ensuring that any improvement of \toolName{} over these baselines can be cleanly attributed to the semantic dependency reconstruction in Stages~1 and~2. Table~\ref{tbl:rq1} presents the reduction results on C and Java bug-triggering programs. For each tool, we report the number of query script invocations~(Q), reduction time in seconds~(T), and the reduced program size in tokens~(R). Each query script invocation validates whether the current reduced program still triggers the target compiler bug. We also report the percentage of tokens produced by \toolName{} relative to those produced by the baselines~(C(\%) w.r.t.).

\textbf{Effectiveness}. \toolName{} consistently outperforms all three syntax-guided baselines in reduction effectiveness. Compared to Perses, \toolName{} achieves a smaller reduced size on 28 out of 28 bug-triggering programs, with a mean further reduction of 51.9\% (154 versus 320 tokens). Compared to WDD and CDD, \toolName{} achieves a smaller reduced size on 24 out of 28 bug-triggering programs, with mean further reduction of 14.9\% and 19.8\% respectively. The gains are most pronounced on complex GCC bugs: gcc-65383 is reduced to 122 tokens by \toolName{}, versus 384, 144, and 156 tokens by Perses, WDD, and CDD respectively (up to 3.1$\times$ smaller). \toolName{} also achieves the smallest reduced size on 11 out of 12 Java bug-triggering programs.
The one notable exception is cf-691, where WDD achieves 95 tokens while \toolName{} produces 454. We manually inspected this case and attributed this to \toolName{}'s default-value reconstruction strategy: replacing deleted references with default values likely destroys the type annotations that trigger this Checker Framework bug, leaving \toolName{} at a larger local minimum. This case suggests that default-value reconstruction can be counterproductive when the bug-triggering property is sensitive to the exact values or types of reconstructed expressions, motivating the same-semantics replacement strategy discussed in Section~\ref{sec:discussion}.

\textbf{Efficiency.} The efficiency results are more nuanced. \toolName{} is slower than Perses on most C bug-triggering programs (mean 1,448 versus 938 seconds, 54.4\% overhead), due to the upfront cost of semantic analysis and dependency graph construction. However, this trend reverses on Java bug-triggering programs, where \toolName{} is faster on 7 out of 12 bugs. Although \toolName{} invokes the query script more frequently than the baselines in many cases, its reduction time is nonetheless lower on these programs. We attributed this to the fact that \toolName{}'s semantic reduction produces smaller intermediate programs earlier in the process --- by deleting large semantic nodes in early rounds, the remaining programs are cheaper to compile and evaluate. As a result, each individual query invocation takes less wall-clock time, compensating for the higher query count. Compared to WDD and CDD, \toolName{} achieves faster reduction on 17 and 18 out of 28 bug-triggering programs respectively, with consistent efficiency advantages on Java programs where semantic dependencies cause the baselines to generate large numbers of invalid intermediate programs that are rejected without making progress.

\paragraph{\textbf{Summary.}} \toolName{} achieves mean size reductions of 51.9\%, 14.9\%, and 19.8\% over Perses, WDD, and CDD respectively, at a moderate efficiency cost on large C programs that is substantially offset on Java bug-triggering programs. These results confirm that semantic dependency reconstruction provides reduction opportunities that purely syntax-guided approaches cannot exploit.

%% file: rq2.tex
\subsection{RQ2: Comparison between \toolName{} and SOTA reducers that incorporate language-specific transformations}
\label{subsec:rq2}

\newcommand{\mybesta}[1]{\cellcolor{green!20}{#1}}

\begin{table*}[htbp]
\centering
\tiny
\caption{Reduction results on C bug-triggering programs compared with reducers that incorporate language-specific transformations
tools. Q = number of query invocations, T = time (seconds), R = reduced size (tokens).  $C = \frac{R_{\text{\toolName{}}} - R_{\text{Baseline}}}{R_{\text{Baseline}}} \times 100\%$ (negative values indicate \toolName{} produces smaller outputs than the baseline).
Best reduced size per bug is highlighted in \colorbox{green!40}{green}.
Second best reduced size is highlighted in \colorbox{green!20}{light green}.
Best time per bug is highlighted in \colorbox{blue!20}{blue}.}
\label{tbl:rq2}
\resizebox{\textwidth}{!}{%
\begin{tabular}{l r | rrr | rrr | rrr | rr}
\toprule
\multirow{2}{*}{\textbf{Bug ID}} &
\multirow{2}{*}{\textbf{Original}} &
\multicolumn{3}{c|}{\textbf{\toolName{}\textsubscript{Perses}}} &
\multicolumn{3}{c|}{\textbf{CReduce~\cite{regehr2012test}}} &
\multicolumn{3}{c|}{\textbf{Latra~\cite{xu2025latra}}} &
\multicolumn{2}{c}{\textbf{C(\%) w.r.t.}} \\
\cmidrule(lr){3-5} \cmidrule(lr){6-8} \cmidrule(lr){9-11}
\cmidrule(lr){12-13}
              &       & Q     & T              & R              & Q     & T     & R            & Q     & T              & R             & Creduce  & Latra    \\ \midrule
\clang{22382} & 21068 & 3459  & \mybestt{118}  & \mybest{95}    & 13857 & 1017  & 110          & 1957  & 126            & \mybesta{109} & -13.6\%  & -12.8\%  \\
\clang{22704} & 93032 & 2888  & \mybestt{223}  & \mybesta{63}   & 10277 & 1111  & \mybest{59}  & 1917  & 273            & \mybest{59}   & 6.8\%    & 6.8\%    \\
\clang{23353} & 30196 & 3763  & \mybestt{236}  & \mybest{72}    & 9297  & 982   & \mybesta{73} & 4932  & 427            & 135           & -1.4\%   & -46.7\%  \\
\clang{25900} & 78960 & 3266  & \mybestt{279}  & \mybesta{107}  & 16775 & 1566  & \mybest{88}  & 4512  & 462            & 148           & 21.6\%   & -27.7\%  \\
\clang{26760} & 32958 & 2612  & \mybestt{151}  & \mybest{50}    & 15965 & 1274  & 65           & 2642  & 217            & \mybesta{59}  & -23.1\%  & -15.3\%  \\
\clang{27747} & 19915 & 2344  & \mybestt{108}  & 115            & 16656 & 1198  & \mybest{106} & 2690  & 181            & \mybesta{107} & 8.5\%    & 7.5\%    \\ \midrule 
\gcc{59903}   & 57581 & 7127  & \mybestt{540}  & \mybest{328}   & 41732 & 3892  & \mybesta{329}& 11377 & 1150           & 398           & -0.3\%   & -17.6\%  \\
\gcc{60116}   & 75224 & 10944 & 1765           & 433            & 58335 & 5248  & \mybesta{394}& 11656 & \mybestt{1713} & \mybest{366}  & 9.9\%    & 18.3\%   \\
\gcc{61383}   & 32449 & 7777  & \mybestt{2503} & 258            & 34140 & 2951  & \mybest{161} & 7717  & 4595           & \mybesta{228} & 60.2\%   & 13.2\%   \\
\gcc{61917}   & 85359 & 8307  & 1698           & \mybesta{127}  & 46557 & 7853  & 329          & 3837  & \mybestt{390}  & \mybest{87}   & -61.4\%  & 46.0\%   \\
\gcc{64990}   & 16623 & 4530  & \mybestt{2042} & \mybesta{163}  & 41382 & 6587  & 243          & 3814  & 2313           & \mybest{64}   & -32.9\%  & 154.7\%  \\
\gcc{65383}   & 43942 & 12945 & \mybestt{3098} & \mybesta{122}  & 41382 & 20184 & 142          & 2542  & 4796           & \mybest{116}  & -14.1\%  & 5.2\%    \\
\gcc{66186}   & 47481 & 5782  & 2347           & \mybesta{236}  & 36294 & 5430  & 280          & 4526  & \mybestt{1816} & \mybest{177}  & -15.7\%  & 33.3\%   \\
\gcc{70127}   & 12358 & 4228  & 4186           & \mybest{107}   & 36482 & 7578  & 277          & 4567  & \mybestt{2466} & \mybesta{109} & -61.4\%  & -1.8\%   \\
\gcc{70586}   & 14626 & 10675 & \mybestt{6145} & \mybesta{346}  & 70493 & 8477  & 417          & 13004 & 6166           & \mybest{301}  & -17.0\%  & 15.0\%   \\
\gcc{71626}   & 6133  & 262   & \mybestt{12}   & 46             & 1832  & 174   & \mybesta{45} & 167   & 13             & \mybest{42}   & 2.2\%    & 9.5\%    \\\midrule
Median        & 15625 & 4379  & 1119           & \mybesta{119}  & 35217 & 3422  & 152 & 4175  & \mybestt{806}  & \mybest{113}           & -21.7\%  & 5.3\%    \\
Mean          & 25192 & 5682  & \mybestt{1591} & \mybesta{167}  & 30716 & 4720  & 195 & 5116  & 1694           & \mybest{157}           & -14.4\%  & 6.4\%    \\
\bottomrule
\end{tabular}%
}
\end{table*}

We compared \toolName{} against two state-of-the-art reducers that incorporate language-specific transformations, CReduce~\cite{regehr2012test} and Latra~\cite{xu2025latra}. Both incorporate hand-crafted, language-specific transformation rules that syntax-guided reducers cannot express, and represent the strongest available points of comparison for C. We restricted this comparison to C because no comparable language-specific reducer exists for Java. Detailed descriptions of each baseline are provided in Section~\ref{sec:related}. Table~\ref{tbl:rq2} presents the reduction results on the 16 C bug-triggering programs.



\textbf{Effectiveness.} Despite having no language-specific transformation rules, \toolName{} achieves the best reduced size on 5 bug-triggering programs (clang-22382, clang-23353, clang-26760, gcc-59903, and gcc-70127) and the second-best on 7 bug-triggering programs (clang-22704, clang-25900, gcc-61917, gcc-64990, gcc-65383, gcc-66186, and gcc-70586), placing first or second on 12 out of 16 bug-triggering programs overall. The mean reduced size of \toolName{} is 167 tokens, situated between CReduce (195 tokens) and Latra (157 tokens) --- a 14.4\% improvement over CReduce and statistically comparable to Latra. This is a notable result given that CReduce relies on more than 5,685 lines of hand-crafted C transformation code and Latra on 27 manually written rules, while \toolName{} requires neither: its advantage derives entirely from language-agnostic semantic dependency reconstruction, which enables deletions that syntax-guided and template-based approaches cannot perform due to unresolved semantic dependencies.

\textbf{Efficiency.} \toolName{} achieves the best reduction time on 12 out of 16 bug-triggering programs, completing in a mean of 1,591 seconds --- 3.0$\times$ faster than CReduce (4,720 seconds) and 1.1$\times$ faster than Latra (1,694 seconds). The efficiency advantage over CReduce is particularly pronounced: CReduce's exhaustive semantic transformations require an average of 30,716 query invocations versus 5,682 for \toolName{} (an 5.4$\times$ reduction), and on individual bug-triggering programs the wall-clock gap can be dramatic --- on clang-22382, \toolName{} completes in 118 seconds while CReduce requires 1,017 seconds.

\paragraph{\textbf{Summary.}} \toolName{} matches or exceeds the effectiveness of CReduce and Latra on the majority of bug-triggering programs while substantially reducing reduction time and query invocations. These results suggest that, on programs where semantic coherence is the primary bottleneck, language-agnostic dependency reconstruction can replace much of the hand-crafted rules in reducers that incorporate language-specific transformations  without sacrificing reduction quality.

%% file: rq3.tex
\subsection{RQ3: Ablation Study of Dependency Reconstruction}
\label{subsec:rq3}

\begin{figure}[htbp]
    \centering
    \begin{minipage}[t]{0.3\textwidth}
        \begin{subfigure}[t]{\linewidth}
            \centering
            \includegraphics[width=\linewidth]{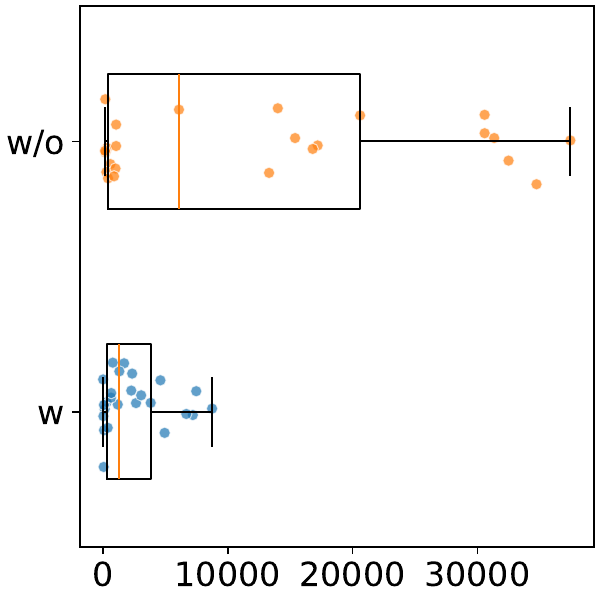}
            \begin{scriptsize}
            \begin{tabular*}{\linewidth}{@{\extracolsep{\fill}} l r r r}
                \toprule
                     & Mean  & Median & STD   \\
                \midrule
                w/o  & 12936 & 13314  & 13237 \\
                w    & 2558  & 1317   & 2717  \\
                \bottomrule
            \end{tabular*}
            \end{scriptsize}
            \caption{Number of query invocations}
            \label{fig:rq2-p}
        \end{subfigure}
    \end{minipage}
    \hfill
    \begin{minipage}[t]{0.3\textwidth}
        \begin{subfigure}[t]{\linewidth}
            \centering
            \includegraphics[width=\linewidth]{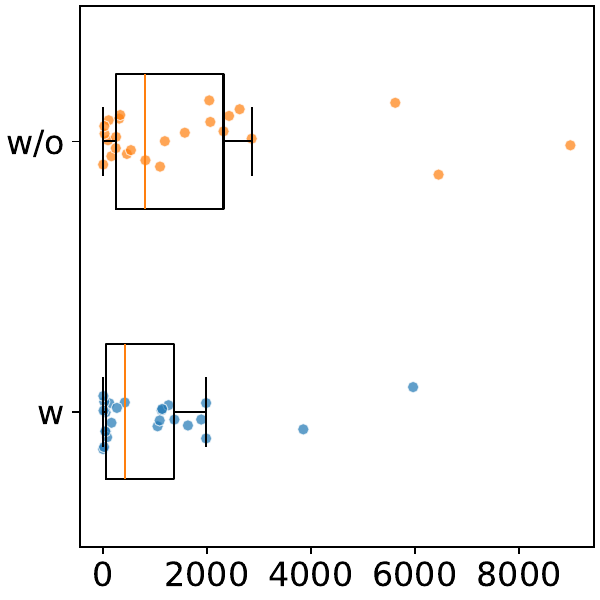}
            \begin{scriptsize}
            \begin{tabular*}{\linewidth}{@{\extracolsep{\fill}} l r r r}
                \toprule
                     & Mean & Median & STD  \\
                \midrule
                w/o  & 2682 & 1100   & 5610 \\
                w    & 1107 & 1052   & 1396 \\
                \bottomrule
            \end{tabular*}
            \end{scriptsize}
            \caption{Reduction time (seconds)}
            \label{fig:rq2-time}
        \end{subfigure}
    \end{minipage}
    \hfill
    \begin{minipage}[t]{0.3\textwidth}
        \begin{subfigure}[t]{\linewidth}
            \centering
            \includegraphics[width=\linewidth]{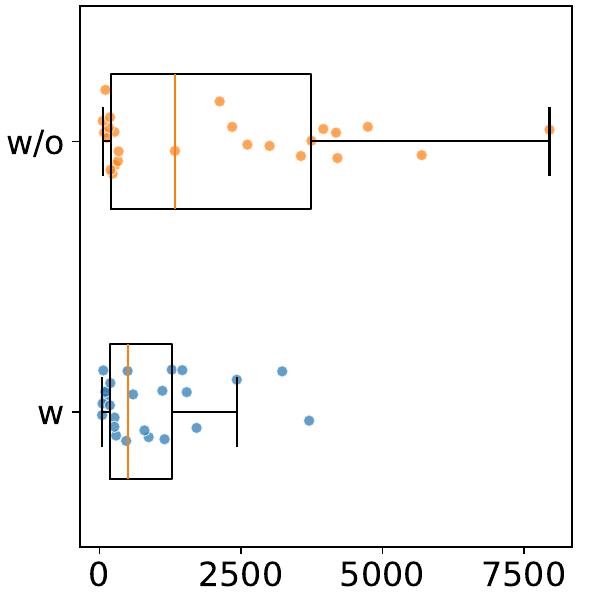}
            \begin{scriptsize}
            \begin{tabular*}{\linewidth}{@{\extracolsep{\fill}} l r r r}
                \toprule
                     & Mean & Median & STD   \\
                \midrule
                w/o  & 2206 & 1335   & 2323  \\
                w    & 990  & 595    & 1076  \\
                \bottomrule
            \end{tabular*}
            \end{scriptsize}
            \caption{Reduced size (tokens)}
            \label{fig:rq2-size}
        \end{subfigure}
    \end{minipage}
    \caption{Ablation study of dependency reconstruction across all 28 bug-triggering programs (w: with dependency reconstruction; w/o: without reconstruction).}
    \label{fig:rq3-result}
\end{figure}

As dependency reconstruction in Stage~2 is the core module of \toolName{}, we conducted an ablation study to evaluate its contribution to overall reduction performance. Specifically, we measured program reduction effectiveness and efficiency with and without dependency reconstruction, using only Stages~1 and~2 and excluding the syntax-guided backend of Stage~3. The reason for this choice is that isolating Stages~1 and~2 allows us to directly attribute any observed difference in effectiveness and efficiency to dependency reconstruction alone. If Stage~3 were included, its syntactic reduction would mask the individual contribution of dependency reconstruction, as improvements could be attributed to the interaction between the semantic and syntactic stages rather than to dependency reconstruction itself. Figure~\ref{fig:rq3-result} reports results on 25 of our 28 bug-triggering programs. We exclude \clang{22704} because, in the \emph{without reconstruction} configuration, the reduction did not converge before our machine exhausted memory. \gcc{70217} and \cf{691} are omitted from the figure as their reduction times in the \emph{without reconstruction} configuration (8,985 and 28,602 seconds respectively) deviate substantially from the rest of the dataset and would distort the visualization. 

\textbf{Efficiency.} Without reconstruction, reduction requires on average 12,936 query invocations (median 13,314) and 2,682 seconds (median 1,100). With reconstruction, query invocations drop to an average of 2,558 (median 1,317) and reduction time decreases to an average of 1,107 seconds (median 1,052) --- representing a 80.2\% reduction in query invocations and a 58.7\% reduction in time. A Mann--Whitney U test confirms that the difference is statistically significant for query invocations ($U = 232.00$, $p = 0.011$) and marginally significant for reduction time ($U = 286.00$, $p = 0.087$). These gains confirm that dependency reconstruction substantially improves efficiency by producing semantically valid intermediate programs that the oracle accepts, thereby avoiding the large numbers of unnecessary query invocations caused by semantically broken intermediates that syntax-guided reduction alone would generate.
 
\textbf{Effectiveness.} Dependency reconstruction also yields a substantial improvement in final reduced size, with the mean token count decreasing from 2,206 to 990 tokens (55.1\% reduction) and the median from 1,335 to 595 tokens (55.4\% reduction). A Mann--Whitney U test confirms that this difference is marginally significant ($U = 272.00$, $p = 0.055$). This confirms that dependency reconstruction not only prevents wasted query invocations, but also unlocks reduction opportunities.

\paragraph{\textbf{Summary.}} Dependency reconstruction reduces query invocations by 80.2\%, reduction time by 58.7\%, and final token count by 55.1\% on average, confirming that it is a central and highly effective component of \toolName{}. 
The consistent gain across both efficiency and effectiveness indicates that semantic coherence breakage is a fundamental bottleneck in language-agnostic program reduction --- one that cannot be addressed by refining syntax-guided search alone.

%% file: discussion.tex
\section{Discussion}
\label{sec:discussion}

\textbf{Applying \methodName{} to Intermediate Representations (IRs).} We believe that \toolName{} may be especially promising for IRs such as LLVM bitcode.
In several bugs, such as \clang{22382}, we observed that developers provided reduced LLVM IR programs rather than reduced C source code, suggesting that IR-level reduction is already a practical option when source-level reduction is insufficient. We hypothesize that applying \toolName{} to LLVM IR could yield significantly higher efficiency than to source code for two reasons. First, IRs are typically designed with explicit representations of semantic dependencies (e.g., LLVM's use-def chains and basic-block structure are part of the IR specification itself), so Stage~1 of \toolName{} can be replaced with a direct translation from IR to dependency graph rather than requiring static analysis. Second, the dependency reconstruction rules required for an IR are typically smaller in number than for a source language, since IRs use a smaller set of constructs and a more regular type system. This is particularly relevant for reduction of bug-triggering inputs produced by fuzzers, which often emit programs in IR form directly. A full evaluation of \toolName{} on IRs is left for future work.

\textbf{Aligning Semantics-aware Reduction with Required Program Properties.}
We believe that when introducing semantics-aware reduction, it is essential to consider the connection between the program properties that need to be preserved and the program semantics. For instance, the reduction methods required for programs that trigger compiler crashes differ significantly from those that cause miscompilations. The former does not require the intermediate reduced programs to be executable, whereas the latter necessitates a certain degree of executability. Note that we do not claim full executability here, as miscompilations can manifest in two ways: a correctly compiled program encountering a runtime error or incorrect output upon completion; both scenarios require specific analysis. Designing reconstruction operators tailored to the semantics of each property could yield more effective reduction than the uniform default-value strategy used by \toolName{}, and is a promising direction for future work.

\textbf{Evaluating Reduction Effectiveness Beyond Token Count.}
Following prior work~\cite{sun2018perses, zhou2024wdd, perses-cdd}, we evaluate reduction effectiveness using token count, which is the standard metric in the program reduction literature. However, we have found that programs with a similar number of tokens can still exhibit significant differences in semantic complexity or readability. In some cases, a program with fewer tokens may actually be more complex than one with more tokens. For example, consider a continuous nested call \texttt{foo(bar(foo()))} versus a sequence of calls introducing temporary variables, such as \texttt{int f = foo(); int b = bar(f); foo(b);}. Although the former has fewer tokens than the latter, its readability is arguably lower. This concern has been discussed in recent work~\cite{perses-sfc}, which proposes structure-form conversion as one approach to producing more readable reduced programs. Developing reduction metrics that better reflect human debugging effort is an open problem in the field, and one to which we believe \toolName{} could contribute, since its dependency-graph-based design enables structurally-aware manipulation of the reduced program rather than only size-based optimization.

\section{Threats to Validity} 
\label{sec:threats}

The primary threats to the validity of our findings are discussed in this section in accordance with established guidelines for empirical software engineering research~\cite{Wohlin2012ESE}.

\textbf{Internal validity.} To ensure fair comparison with baselines, all tools (\toolName{}, Perses, WDD, CDD, CReduce, and Latra) are run from their publicly released implementations on the same machine, under single-thread execution. We acknowledge that further tuning of the baselines' configurations could potentially improve their performance, but we do not perform such tuning to keep the comparison neutral. 

\textbf{External validity.} \toolName{} currently supports C and Java; while we argue in Section~\ref{sec:impl} that the design generalizes to any programming language with a static type system, we have not empirically evaluated this claim. Evaluating \toolName{} on additional languages and intermediate representations remains future work.

\textbf{Construct validity.} We use token count as our primary metric for reduction effectiveness, following prior work~\cite{sun2018perses, zhou2024wdd, perses-cdd}, but token count is an imperfect indicator of human debugging effort, as discussed in Section~\ref{sec:discussion}. Our efficiency metric, wall-clock time, is sensitive to system load and concurrent processes; we mitigate this threat by running all experiments on our machine with single-thread execution.

%% file: related.tex
\section{Related Work}
\label{sec:related}
We discuss language-agnostic reducers, reduction with language-specific transformations, and approaches for avoiding or repairing invalid intermediate programs, and highlight how our work differs from these existing approaches.

\subsection{Language-agnostic Reduction}

Delta Debugging (DDMin)~\cite{ddmin} reduces failure-inducing programs by iteratively deleting code fragments while preserving the failure-triggering behavior. However, text-based DDMin often generates many invalid intermediate programs when applied to structured inputs such as source code. Hierarchical Delta Debugging (HDD)~\cite{misherghi2006hdd} addresses this issue by performing reductions over Abstract Syntax Trees (ASTs) rather than raw text, ensuring that intermediate candidates remain syntactically valid. Perses~\cite{sun2018perses} further extends this idea by integrating grammars with delta debugging and applying grammar-preserving transformations to systematically minimize programs without violating syntactic constraints.

Building on Perses, Vulcan~\cite{xu2023vulcan} goes beyond 1-minimality through three additional operations: smaller-structure replacement, identifier replacement, and subtree replacement. T-Rec~\cite{perses-trec} complements tree-level reductions with fine-grained lexical reduction that exploits sub-token reduction opportunities. PPR~\cite{zhang2023ppr} extends program reduction to a pairwise setting by simultaneously minimizing both a bug-triggering program and a passing variant while minimizing their differences, thereby highlighting the bug-inducing change.

Another line of work focuses on improving the underlying delta debugging algorithm. ProbDD~\cite{pdd} replaces the traditional delta debugging strategy with a probabilistic model that estimates the likelihood of each element being failure-relevant and selects deletion subsets that maximize the expected number of removable elements per attempt. CDD~\cite{perses-cdd} simplifies ProbDD by showing that its probabilities can be analytically precomputed as counters, achieving comparable performance with substantially lower complexity. WDD~\cite{zhou2024wdd} augments Perses by assigning weights to syntax elements based on token count, prioritizing the deletion of smaller and more likely irrelevant elements.

\subsection{Reduction with Language-specific Transformations}
C-Reduce~\cite{regehr2012test} is a language-specific reducer for C programs that applies more than 30 compiler-like source-to-source transformations, including simultaneously removing function parameters and their corresponding call-site arguments, inlining function bodies, and other semantics-aware simplifications. ReduKtor~\cite{stepanov2019reduktor} adopts a hybrid approach for Kotlin compiler testing by combining program slicing, hierarchical delta debugging, and language-specific transformations. However, its reliance on manually engineered Kotlin-specific transformations limits portability to other languages, reflecting the trade-off between effectiveness and generality that \toolName{} aims to address.

More recent work has explored ways to reduce the engineering burden of language-specific reducers. LPR~\cite{zhang2024lpr} combines the language generality of reducers such as Perses with language-specific semantics inferred by Large Language Models (LLMs), demonstrating effectiveness on C, Rust, and JavaScript programs. Latra~\cite{xu2025latra} bridges the gap between language-agnostic and language-specific reducers by augmenting a language-agnostic reducer with user-defined transformations expressed as match-rewrite template pairs in a domain-specific language. For C programs, Latra implements 27 transformation rules and achieves results statistically comparable to C-Reduce while requiring up to 34$\times$ fewer lines of code.

\subsection{Avoiding Invalid Intermediate Programs}
Several studies address the same fundamental challenge targeted by \toolName{}: avoiding invalid intermediate programs that significantly increase reduction cost.

J-Reduce~\cite{kalhauge2019binary,kalhauge2021logical} uses propositional logic to specify dependencies and formulates Java bytecode reduction as a logical satisfiability problem, ensuring that intermediate programs remain semantically sound. However, J-Reduce derives dependencies from Java bytecode semantics and is therefore tightly coupled to a specific language, whereas \toolName{} employs a language-agnostic dependency model. In addition, \toolName{} proactively reconstructs broken dependencies during reduction, while J-Reduce primarily prevents invalid reductions through dependency constraints. Gharachorlu and Sumner~\cite{gharachorlu2021model} train a machine learning model on syntactic features to predict which deletion candidates are likely to remain semantically valid. GReduce~\cite{ren2025greduce} takes a different direction: instead of reducing the generated program itself, it minimizes the execution trace of the \emph{generator} that produces the program, thereby ensuring program validity inherently. Specimin~\cite{specimin}, the source of the Checker Framework benchmarks used in our evaluation, performs \emph{static} program reduction by exploiting the modularity of type rules to compute a slice that preserves the typechecker's behavior, without running the typechecker on each candidate. Specimin is more suitable for bugs where the error messages are more explicit, while \toolName{} is a dynamic reducer that targets any property a query script can express, including compiler crashes and miscompilations.

\subsection{Conclusive Summary}
\toolName{} strikes a balance between fully language-agnostic approaches and reduction techniques based on language-specific transformations. While remaining language-agnostic, it introduces a lightweight semantic layer that repairs broken dependencies during reduction. This enables \toolName{} to recover much of the semantic coherence achieved by language-specific reducers without requiring extensive per-language engineering. In addition, \toolName{} complements existing approaches that aim to \emph{predict} or \emph{prevent} invalid intermediate programs by instead proactively \emph{repairing} invalid intermediates through dependency reconstruction.



%% file: conclusion.tex
\section{Conclusions}
\label{sec:conclusion}

This paper presents \toolName{}, a novel program reduction approach that enhances syntax-guided program reduction with dependency reconstruction. By repairing broken dependencies through default-value reconstruction, \toolName{} preserves the semantic validity of intermediate programs and unlocks reduction opportunities that purely syntax-guided approaches cannot exploit. Our evaluation on real-world bug-triggering C and Java programs shows that (1) \toolName{} produces significantly smaller reduced programs than state-of-the-art syntax-guided reducers with competitive efficiency, and (2) \toolName{} achieves results comparable to reducers that incorporate language-specific transformations without using any such transformations itself, with improved efficiency overall.
An ablation study further confirms that dependency reconstruction is the key contributor to \toolName{}'s performance, reducing query invocations by 80.2\% and final token count by over 55.1\% on average.

In the next step, we plan to extend \toolName{} along two directions. First, we will broaden its applicability by supporting additional languages and intermediate representations such as LLVM IR, where semantic dependencies are already explicit and reduction may be more efficient. Second, we will explore more sophisticated dependency reconstruction strategies, including same-semantics replacement that preserves program behavior, to address the limitation observed on bugs sensitive to specific values or types.

%% file: ref.bib
@String{Springer = "Springer-Verlag" }

@inproceedings{stepanov2019reduktor,
  title = {ReduKtor: How We Stopped Worrying About Bugs in Kotlin Compiler},
  booktitle = {Proceedings of the 34th IEEE/ACM International Conference on Automated Software Engineering (ASE)},
  author = {Stepanov, Daniil and Akhin, Marat and Belyaev, Mikhail},
  pages = {317--326},
  publisher = {IEEE},
  year={2019}
}

@article{specimin,
  title = {Static {{Program Reduction}} via {{Type-Directed Slicing}}},
  author = {Nguyen, Loi Ngo Duc and Islam, Tahiatul and Wang, Theron and Lenz, Sam and Kellogg, Martin},
  year = 2025,
  journal = {Proceedings of the ACM on Software Engineering},
  volume = {2},
  number = {ISSTA},
  pages = {2068--2090}
}

@misc{psi,
  author={JetBrains},
  title={Program Structure Interface},
  year={2024},
  url={https://plugins.jetbrains.com/docs/intellij/psi.html}
}

@inproceedings{pdd,
  author = {Wang, Guancheng and Shen, Ruobing and Chen, Junjie and Xiong, Yingfei and Zhang, Lu},
  title = {Probabilistic Delta Debugging},
  year = {2021},
  booktitle = {Proceedings of the 29th ACM Joint Meeting on European Software Engineering Conference and Symposium on the Foundations of Software Engineering (ESEC/FSE)},
  organization={ACM},
  pages = {881--892},
}

@misc{trevino2019llvm-reduce,
  author       = {Diego Trevi\~{n}o Ferrer},
  title        = {{LLVM-Reduce} for testcase reduction},
  year         = {2019},
  note         = {\url{https://llvm.org/devmtg/2019-10/talk-abstracts.html\#tech22}},
}

@misc{gcc-bugs-guide,
  author       = {GNU Compiler Collection Contributors},
  title        = {Reporting Bugs},
  year         = {2026},
  note         = {\url{https://gcc.gnu.org/bugs/}},
}

@misc{llvm-how-to-submit-bug,
  author       = {{LLVM Project}},
  title        = {How to Submit a Bug Report},
  year         = {2026},
  note         = {\url{https://llvm.org/docs/HowToSubmitABug.html}},
}

@inproceedings{chaliasos2022finding,
  title={Finding typing compiler bugs},
  author={Chaliasos, Stefanos and Sotiropoulos, Thodoris and Spinellis, Diomidis and Gervais, Arthur and Livshits, Benjamin and Mitropoulos, Dimitris},
  booktitle={Proceedings of the 43rd ACM SIGPLAN International Conference on Programming Language Design and Implementation (PLDI)},
  pages={183--198},
  organization={ACM},
  year={2022}
}

@inproceedings{xu2025latra,
  title={Latra: A Template-Based Language-Agnostic Transformation Framework for Effective Program Reduction},
  author={Xu, Zhenyang and Wang, Yiran and Tian, Yongqiang and Zhang, Mengxiao and Sun, Chengnian},
  booktitle={Proceedings of the 40th IEEE/ACM International Conference on Automated Software Engineering (ASE)},
  pages={2274--2285},
  organization={IEEE},
  year={2025}
}

@inproceedings{kalhauge2019binary,
  title={Binary reduction of dependency graphs},
  author={Kalhauge, Christian Gram and Palsberg, Jens},
  booktitle={Proceedings of the 27th ACM Joint Meeting on European Software Engineering Conference and Symposium on the Foundations of Software Engineering (ESEC/FSE)},
  pages={556--566},
  organization={ACM},
  year={2019}
}

@inproceedings{regehr2012test,
  title={Test-case reduction for C compiler bugs},
  author={Regehr, John and Chen, Yang and Cuoq, Pascal and Eide, Eric and Ellison, Chucky and Yang, Xuejun},
  booktitle={Proceedings of the 33rd ACM SIGPLAN Conference on Programming Language Design and Implementation (PLDI)},
  pages={335--346},
  organization={ACM},
  year={2012}
}

@article{feng2025finding,
  title={Finding Compiler Bugs through Cross-Language Code Generator and Differential Testing},
  author={Feng, Qiong and Ma, Xiaotian and Feng, Ziyuan and Akhin, Marat and Song, Wei and Liang, Peng},
  journal={Proceedings of the ACM on Programming Languages},
  volume={9},
  number={OOPSLA2},
  pages={2843--2869},
  year={2025},
  publisher={ACM}
}

@article{ddmin,
  title={Simplifying and isolating failure-inducing input},
  author={Zeller, Andreas and Hildebrandt, Ralf},
  journal={IEEE Transactions on Software Engineering},
  volume={28},
  number={2},
  pages={183--200},
  year={2002},
  publisher={IEEE}
}

@inproceedings{misherghi2006hdd,
  title={HDD: Hierarchical delta debugging},
  author={Misherghi, Ghassan and Su, Zhendong},
  booktitle={Proceedings of the 28th International Conference on Software Engineering (ICSE)},
  pages={142--151},
  organization={ACM},
  year={2006}
}

@inproceedings{kalhauge2021logical,
  title={Logical bytecode reduction},
  author={Kalhauge, Christian Gram and Palsberg, Jens},
  booktitle={Proceedings of the 42nd ACM SIGPLAN International Conference on Programming Language Design and Implementation (PLDI)},
  pages={1003--1016},
  organization={ACM},
  year={2021}
}

@inproceedings{sun2018perses,
  title={Perses: Syntax-guided program reduction},
  author={Sun, Chengnian and Li, Yuanbo and Zhang, Qirun and Gu, Tianxiao and Su, Zhendong},
  booktitle={Proceedings of the 40th International Conference on Software Engineering (ICSE)},
  pages={361--371},
  organization={ACM},
  year={2018}
}

@inproceedings{zhang2023ppr,
  title={PPR: Pairwise Program Reduction},
  author={Zhang, Mengxiao and Xu, Zhenyang and Tian, Yongqiang and Jiang, Yu and Sun, Chengnian},
  booktitle={Proceedings of the 31st ACM Joint European Software Engineering Conference and Symposium on the Foundations of Software Engineering (ESEC/FSE)},
  pages={338--349},
  organization={ACM},
  year={2023}
}

@inproceedings{zhang2024lpr,
  title={LPR: Large language models-aided program reduction},
  author={Zhang, Mengxiao and Tian, Yongqiang and Xu, Zhenyang and Dong, Yiwen and Tan, Shin Hwei and Sun, Chengnian},
  booktitle={Proceedings of the 33rd ACM SIGSOFT International Symposium on Software Testing and Analysis (ISSTA)},
  pages={261--273},
  organization={ACM},
  year={2024}
}

@inproceedings{zhou2024wdd,
  author = {Zhou, Xintong and Xu, Zhenyang and Zhang, Mengxiao and Tian, Yongqiang and Sun, Chengnian},
  booktitle={Proceedings of the 47th IEEE/ACM International Conference on Software Engineering (ICSE)},
  title = {WDD: Weighted Delta Debugging},
  year = {2025},
  organization={IEEE},
  pages = {1592--1603}
}

@article{perses-sfc,
  title={Boosting Program Reduction with the Missing Piece of Syntax-Guided Transformation},
  author={Xu, Zhenyang and Tian, Yongqiang and Zhang, Mengxiao and Sun, Chengnian},
  journal={Proceedings of the ACM on Programming Languages},
  year={2025},
  volume={9},
  number={OOPSLA2},
  publisher = {ACM},
  pages={86--112}
}

@inproceedings{perses-cdd,
  title={Toward a Better Understanding of Probabilistic Delta Debugging},
  author={Zhang, Mengxiao and Xu, Zhenyang and Tian, Yongqiang and Cheng, Xinru and Sun, Chengnian},
  booktitle={Proceedings of the 47th International Conference on Software Engineering (ICSE)},
  year={2025},
  organization={ACM},
  pages={2024--2035}
}

@article{perses-trec,
  title={T-rec: Fine-grained language-agnostic program reduction guided by lexical syntax},
  author={Xu, Zhenyang and Tian, Yongqiang and Zhang, Mengxiao and Zhang, Jiarui and Liu, Puzhuo and Jiang, Yu and Sun, Chengnian},
  journal={ACM Transactions on Software Engineering and Methodology},
  volume={34},
  number={2},
  pages={1--31},
  year={2025},
  publisher={ACM}
}

@inproceedings{zhang2021sanrazor,
  title={SANRAZOR: Reducing Redundant Sanitizer Checks in C/C++ Programs},
  author={Zhang, Jiang and Wang, Shuai and Rigger, Manuel and He, Pinjia and Su, Zhendong},
  booktitle={Proceedings of the 15th USENIX Symposium on Operating Systems Design and Implementation (OSDI)},
  pages={479--494},
  organization={USENIX Association},
  year={2021}
}

@book{sedgewick2011algorithms,
  title={Algorithms},
  author={Sedgewick, Robert and Wayne, Kevin},
  year={2011},
  publisher={Addison-Wesley Professional}
}

@article{xu2023vulcan,
  title={Pushing the limit of 1-minimality of language-agnostic program reduction},
  author={Xu, Zhenyang and Tian, Yongqiang and Zhang, Mengxiao and Zhao, Gaosen and Jiang, Yu and Sun, Chengnian},
  journal={Proceedings of the ACM on Programming Languages},
  volume={7},
  number={OOPSLA1},
  pages={636--664},
  year={2023},
  publisher={ACM}
}

@inproceedings{gharachorlu2021model,
  title={Leveraging models to reduce test cases in software repositories},
  author={Gharachorlu, Golnaz and Sumner, Nick},
  booktitle={Proceedings of the 18th IEEE/ACM International Conference on Mining Software Repositories (MSR)},
  pages={230--241},
  year={2021},
  organization={IEEE}
}

@article{ren2025greduce,
  title={Validity-Preserving Delta Debugging via Generator Trace Reduction},
  author={Ren, Luyao and Zhang, Xing and Hua, Ziyue and Jiang, Yanyan and He, Xiao and Xiong, Yingfei and Xie, Tao},
  journal={ACM Transactions on Software Engineering and Methodology},
  volume={34},
  number={3},
  pages={1--33},
  year={2025},
  publisher={ACM}
}

@inproceedings{sunfinding2016,
	title = {Finding compiler bugs via live code mutation},
	pages = {849--863},
	booktitle = {Proceedings of the 31st ACM SIGPLAN International Conference on Object-Oriented Programming, Systems, Languages, and Applications (OOPSLA)},
    organization={ACM},
    year={2016},
	author = {Sun, Chengnian and Le, Vu and Su, Zhendong}
}

@article{lidbury-many-core-2015,
  title={Many-core compiler fuzzing},
  author={Lidbury, Christopher and Lascu, Andrei and Chong, Nathan and Donaldson, Alastair F},
  journal={ACM SIGPLAN Notices},
  volume={50},
  number={6},
  pages={65--76},
  year={2015},
  publisher={ACM}
}

@inproceedings{yangfinding2011,
	title = {Finding and understanding bugs in C compilers},
	pages = {283--294},
	booktitle = {Proceedings of the 32nd ACM SIGPLAN Conference on Programming Language Design and Implementation (PLDI)},
	organization = {ACM},
    year={2011},
	author = {Yang, Xuejun and Chen, Yang and Eide, Eric and Regehr, John}
}

@book{Wohlin2012ESE,
  title={Experimentation in Software Engineering},
  author={Wohlin, Claes and Runeson, Per and H{\"o}st, Martin and Ohlsson, Magnus C. and Regnell, Bj{\"o}rn and Wessl{\'e}n, Anders},
  publisher={Springer},
  year={2012}
}
